\documentstyle{article}

\begin{document}
\baselineskip=18pt
\title{Bosonization of quantum sine-Gordon field with boundary }

\author{
 Bo-yu Hou$^{\dagger}$ 
\hskip 0.5truecm Kang-jie Shi$^{\dagger}$ \\
Yan-shen Wang$^{\ddag,\dagger}$ \hskip 0.5truecm 
Wen-li Yang$^{\ddag,\dagger}$ \\
\bigskip\\
$^{\ddag}$ CCAST (World Laboratory), P.O.Box 8730, Beijing 100080, China\\
$^{\dagger}$ Institute of Modern Physics, Northwest 
University, Xian 710069, China
\thanks{Mailing address}}
\date{March 1, 1996}
\maketitle

\begin{abstract}
Boundary operators and boundary ground states in sine-Gordon model with a
 fixed boundary condition are studied using bosonization and 
q-deformed oscillators.We also obtain the form-factors of this model.

\end{abstract}

\section{Introduction}

An integrable two-dimensional field theory possesses an infinite set of mutually commutative
integrals of motion.For the massive theory,this results in the factorizable 
scattering both in bulk case (i.e without a boundary ) and in the case 
with an integrable reflecting boundary (e.g. field theories defined upon the half line ) 
$^{[1,2,3,4]}$ .In the boundary case ,besides the two particle scattering 
S-matrix, there exists boundary reflecting R-matrix,which describes the 
process of particle reflecting at the boundary. The R-matrix must 
be consent with the 
S-matix (namely, R-matrix must satisfy the boundary Yang-Baxter 
equation$^{[2,29,30]}$).The calculation of the exact factorized S-matrix and 
R-matrix may be 
performed by combining the standard requirements of unitarity and crossing 
symmetry together with the symmetry properties of the model $^{[1,2]}$.

In the bulk case,the two particle S-matrix uniquely specifies the structure 
of the space of local operators for an integrable model.Its 
knowledge can be used to calculate off-shell quantities,like correlation 
functions of elementary or composite fields of the integrable models under 
investigation.These quantities can be obtained by considering the 
form-factors of 
 local fields,which are matrix elements of operators between asymptotic 
states$^{[5,6,10]}$.It is shown that the general properties of unitary,
analyticity and locality lead to a system of functional equations on 
form-factors.These equations include S-matrix elements as structure constants 
and are enough for the resconstruction of the form-factors (see 
also [10][11][12]).Therefore, the S-matrix determines the whole fields 
in the bulk case.Obtaining the form-factors, one can in principle obtain 
all the off-shell quantities of the field theory.We next introduce these 
equations for the form-factors of any local operator 
in the massive integrable model without boundary.It is shown that the 
form-factors for any local field O
\begin{eqnarray*}
& & f(\beta_{1},....,\beta_{n})\equiv \langle 0|OZ(\beta_{1})....Z(\beta_{n})|0\rangle
\end{eqnarray*}
are subject to a series 
axioms$^{[10,12]}$ as the following:

1.Riemann-Hilbert problem
\begin{eqnarray}
& &S_{ij}(\beta_{i}-\beta_{j})f(\beta_{1},...,
\beta_{i},...,\beta_{j},...,\beta_{n})=f(\beta_{1},...,\beta_{j},...,
\beta_{i},...,\beta_{n})\nonumber\\
& &f(\beta_{1},...,\beta_{i},...,\beta_{n-1},\beta_{n}+2\pi i)=
f(\beta_{n},\beta_{1},...,\beta_{i},...,\beta_{n-1})\ \ \ ;
\end{eqnarray}

2.Residue condition
\begin{eqnarray}
& &2\pi iRes_{\beta_{n}=\beta_{n-1}+i\pi}f(\beta_{1},....,\beta_{n-1},
\beta_{n})_{a_{1}....a_{n}}=f(\beta_{1},....,\beta_{n-2})_{a'_{1},....,
a'_{n-2}}C_{a'_{n-1},a_{n}}\nonumber\\
& &\{\delta^{a'_{1}}_{a_{1}}....\delta^{a'_{n-2}}_{a_{n-2}}
\delta^{a'_{n-1}}_{a_{n-1}}-exp(2i\pi\omega(O,\Psi))S_{c_{1},a_{1}}^{a'_{n-1},a'_{1}}(\beta_{n-1}-
\beta_{1})....\nonumber\\
& &\ \ \ \ S_{c_{n-3},a_{n-3}}^{c_{n-4},a'_{n-3}}
(\beta_{n-1}-\beta_{n-3}) S_{a_{n-1},a_{n-2}}^{c_{n-3},a'_{n-2}}(\beta_{n-1}
-\beta_{n-2})\}\ \ \ .
\end{eqnarray}
In Eq.s(1),(2), we use the usual conventions $^{[10,12]}$ : $S(\beta)$ 
is a two particle S-matrix and matrix $C$ is related to the crossing 
symmetry of S-matrix ;the number 
of $\omega (O,\Psi)$ is the mutual locality index for the local operator $O$ 
with respect to the ``elementary field" $\Psi$ $^{[12]}$.Therefore,solving 
the equations (1),(2) plays a very important role in the massive integrable 
model without boundary.There are two ways to solve the 
equations (1),(2).The first one is to directly solve the difference equations 
of q-deformed Knizhnik-Zamolodchikov (KZ) equations$^{[10]}$ 
(q-deformed KZ equations are 
equivalent to equations (1) and (2) ),which can be considered as a 
generalization of BPZ's work in two dimensional conformal field theories 
(CFT) $^{[15]}$.In this way,Smirnov obtained the form-factors of some 
local operators for SU(2)-Invariant Thirring model (SU(2)-ITM),sine-Gordon 
(SG) model and O(3)-Nolinear $\sigma $-model$^{[10,35]}$.
Smirnov and Reshetikhin 
discussed some properties of the Restricted SG model$^{[18]}$.
The second way is by bosonization.In the second way ,following the success of the bosonization in 
CFT $^{[16,17,18,19]}$ ,much attention has 
been focused on the vertex operators of some q-deformed affine Lie algebra 
which satisfy q-deformed KZ equation $^{[21,24,38]}$.Using these 
vertex operators,one 
can obtain the bosonization of massive integrable field models and solvable 
lattice models $^{[23,39,12]}$.This method can be considered as the generalization of 
bosonization in CFT$^{[16,17,19,20]}$ and may be a very powerful and 
efficient method.Using this bosonization method , one can get the 
general integal representations for 
form-factors in SU(2)-IMT,SG and XXZ model$^{[12,25]}$.

Although much progress has been obtained in the massive integrable model 
without boundary (in the bulk case),the progress of massive integrable model with 
integrable boundary condition is still very few.Recently,Ghoshal and 
Zamolodchikov $^{[2]}$ proposed boundary crossing relation to determine the 
scalar factor of the boundary reflecting matrix (R-matrix).
In the boundary case,it is S-matrix and R-matrix that uniquely specify 
the structure of the model under investigation.From the S-matrix and  
R-matrix,Jimbo et al obtained the general integral representations of 
form-factors for XXZ model with an integrable boundary (a fixed boundary 
condition),and found that form-factors for this model satisfied an 
analogues of difference equation$^{[26]}$ which has only a subtle difference 
from the bulk case.Some solution to these 
equations in the XYZ model with boundary was already obtained$^{[26]}$.
In this paper, we use the method of bosonization for 
integrable model to study the SG model with a fixed boundary condition,
which has been proved to be integrable$^{[2,3,4,30,31]}$.It is the extension 
of Lukyanov's work,which only deals with SG model without boundary (in 
the bulk case),to the integrable boundary case.However,this method can also 
be extended to other boundary integrable models$^{[36]}$.We find that in the 
boundary case of massive integrable quantum field theories,the form-factors 
also satisfy some difference equations which are quite similar to the solvable 
lattice model.These functional equations are enough to 
reconstruct the form-factors as in the bulk case.

This papper is organized as follows.In sect.2 we give a brief review 
of SG model without boundary (bulk case) by  Lukyanov 
which will be used in the 
following section.Boundary operators with the bosonic version will be 
studied 
in sect.3,from these boundary operators we obtain the boundary bound ground 
states and find the reflecting matrix for the quantum breathers with these 
bound states.In order 
to calculate the form-factors, the boundary operators with the 
q-deformed oscillators expression is given in sect.4.
Then, the form-factors are represented in the term of general 
integrals in sect.5. Thus in principle we can obtain all the quantities of 
the SG model with an integrable boundary.

\section{SG model in the bulk }
  The  SG field theory in the bluk case is described by the action 
as the following
\begin{eqnarray}
{\cal L}=\int_{-\infty }^{\infty}dx\int_{-\infty}^{\infty}dt \frac{1}{2}(
\partial_{\mu} \varphi )^{2}-\frac{m^{2}}{\beta ^{2}}cos\beta \varphi 
\end{eqnarray}
where $\beta$ is the bare coupling constant.In what follows we shall only use 
the renormalized coupling constant $\xi =\frac{\beta^{2}}
{8\pi ^2-\beta ^{2}}$ and for the simplicity we consider a generic 
$\xi$ . 
It is well-known that this theory is massive and its particle spectrum consists of a kink-antikink
pair ($Z_{-},Z_{+}$) ( with equal masses ) and a number of netural particles 
$B_{n}$ (`` quantum breathers" ) whose number is dependent on renormalized 
coupling constant $\xi$($n<[\frac{1}{\xi}]$)$^{[1,2,27,28]}$.The asymptotic 
particle states are generated by the particle-creation operators 
$Z_{a}(\beta) \ \ (a=\pm)$
\begin{eqnarray}
|Z_{a_{1}}(\beta _{1})Z_{a_{2}}(\beta _{2}).... Z_{a_{N}}(\beta _{N})
\rangle =Z_{a_{1}}(\beta _{1}).... Z_{a_{N}}(\beta _{N})|0\rangle  
\end{eqnarray}
where $\beta_{i} $ is particle's rapidity and operators $Z_{a}$ satisfy the 
Zamolodchikov-Faddeev (ZF) algebra:
\begin{eqnarray}
Z_{i}(\beta _{1})Z_{j}(\beta _{2})=S_{ij}^{kl}(\beta _{1}-\beta _{2})Z_{l}
(\beta _{2})Z_{k}(\beta_{1})  \ \ \ .
\end{eqnarray}                     
Lukyanov $^{[12]}$ constructed two kinds of operators : asymptotic operators 
$Z_{i}(\beta )$ and local operators $Z_{i}^{'}(\beta)$
--- using free bosons and q-deformed oscillators,which behave quite similar to 
the type I and type  II operators in the exact solvable lattice model XXZ
$^{[23]}$ .These two kinds of operators satisfy relations (5) and (6)
\begin{eqnarray}
Z_{i}^{'}(\beta _{1})Z_{j}^{'}(\beta _{2})=S_{ij}^{'kl}(\beta _{1}-\beta _{2})
Z_{l}^{'}(\beta _{2})Z_{k}^{'}(\beta _{1}),\ \ \ 
Z_{i}(\beta _{1})Z_{j}^{'}(\beta _{2})=ijtg(\frac{\pi }{4}+i\frac{\beta _{1}
-\beta _{2}}{2})Z_{j}^{'}(\beta _{2})Z_{i}(\beta _{1})\ \ .  
\end{eqnarray}
The S-matrix in (5) and (6) are given as 
\begin{eqnarray*}
& &S_{12}(\beta )=s(\beta )\left( \begin{array}{llll}1&&&\\
&\frac{sh\frac{\beta }{\xi }}{sh\frac{i\pi -\beta }{\xi }}&\frac{sh\frac{i\pi }
{\xi }}{sh\frac{i\pi -\beta}{\xi }}&\\
&\frac{sh\frac{i\pi }{\xi }}{sh\frac{i\pi -\beta }{\xi }}&\frac{sh\frac{\beta }
{\xi }}{sh\frac{i\pi -\beta }{\xi }}&\\
&&&1\\
\end{array}
\right) \ \ , \ \ 
S^{'}_{12}(\beta )=s^{'}(\beta )\left( \begin{array}{llll}1&&&\\
&-\frac{sh\frac{\beta }{\xi +1}}{sh\frac{i\pi +\beta }{\xi +1}}&\frac{sh
\frac{i\pi}{\xi +1}}{sh\frac{i\pi +\beta }{\xi +1}}&\\
&\frac{sh\frac{i\pi}{\xi +1}}{sh\frac{i\pi +\beta}{\xi +1}}&-\frac{sh\frac{
\beta }{\xi +1}}{sh\frac{i\pi +\beta }{\xi +1}}&\\
&&&1\\
\end{array}
\right) ,\\
& &s(\beta )=\frac{\Gamma (\frac{1}{\xi }-\frac{i\beta }{\pi \xi})\Gamma(\frac{i
\beta }{\pi \xi })}{\Gamma (-\frac{i\beta }{\pi \xi })\Gamma (\frac{1}{\xi }+
\frac{i\beta }{\pi \xi })}\prod_{p=1}^{\infty }\frac{R_{p}(\beta)}{R_{p}
(-\beta )} \ \ , \ \ 
R_{p}(\beta)=\frac{\Gamma (\frac{2p}{\xi}+\frac{i\beta}{\pi \xi})\Gamma(1+\frac
{2p}{\xi}+\frac{i\beta}{\pi \xi })}{\Gamma (\frac{2p+1}{\xi }+\frac{i\beta}{\pi
 \xi })\Gamma (1+\frac{2p-1}{\xi}+\frac{i\beta}{\pi \xi})} \ \ ,\\
& &s^{'}(\beta)=\frac{\Gamma(1-\frac{i\beta}{\pi(\xi+1)})\Gamma(\frac{
1}{\xi+1}+\frac{i\beta}{\pi(\xi+1)})}{\Gamma(1+\frac{i\beta}{\pi(\xi+1)})\Gamma(
\frac{1}{\xi+1}-\frac{i\beta}{\pi(\xi+1)})}\prod^{\infty}_{p=1}\frac{R^{'}_{p}
(-\beta)}{R^{'}_{p}(\beta)} \ \ , \ \ \
 R_{p}^{'}(\beta )=\frac{\Gamma(\frac{2p}{\xi+1}+\frac{i\beta}{\pi(\xi+1)})
\Gamma(1+\frac{2p}{\xi+1}+\frac{i\beta}{\pi(\xi+1)})}
{\Gamma(\frac{2p+1}{\xi+1}+\frac{i\beta}{\pi(\xi+1)})\Gamma(1+\frac{2p-1}{\xi+1}
+\frac{i\beta}{\pi(\xi+1)})} \ \ .
\end{eqnarray*} 
The operators $Z_{i}(\beta ),Z^{'}_{j}(\beta )$ can be realized through the bosonic field 
$\phi (\beta )$ and $\phi ^{'}(\beta )$ as
\begin{eqnarray}
& &Z_{+}(\beta)=e^{(-\frac{\beta}{2\xi})}V(\beta)\equiv e^{(-\frac{\beta}{2\xi})}
:e^{i\phi (\beta)}: \ \ , \ \  
Z_{-}(\beta)=e^{(\frac{\beta}{2\xi})}[q^{\frac{1}{2}}\chi V(\beta)-q^{-\frac{1}
{2}}V(\beta)\chi] ,\nonumber\\
& &Z_{+}^{'}(\beta)=e^{(-\frac{\beta}{2(\xi+1)})}V^{'}(\beta)\equiv e^{(-\frac{
\beta}{2(\xi +1)})}:e^{i\phi ^{'}(\beta)}: \ \ ,\ \ 
Z_{-}^{'}(\beta)=e^{(\frac{\beta}{2(\xi +1)})}[q^{'\frac{1}{2}}\chi ^{'}V^{'}
(\beta)-q^{'-\frac{1}{2}}V^{'}(\beta)\chi^{'}],\nonumber\\
& &q=exp\{i\pi \frac{\xi +1}{\xi }\} \ \ \  , \ \ \ q^{'}=exp\{i\pi \frac{\xi }
{\xi +1}\}\ \ \ ,
\end{eqnarray}
where $\phi(\beta)$ and $\phi^{'}(\beta)$ satisfy
\begin{eqnarray*}
& &\langle 0|\phi(\beta _{1})\phi(\beta _{2})|0\rangle =-lng(\beta _{2}-\beta _
{1}),\ \ s(\beta )=\frac{g(-\beta)}{g(\beta)},\ \ 
g(\beta)=\frac{\Gamma(\frac{1}{\xi}+\frac{i\beta}{\pi\xi})}{\Gamma(
\frac{i\beta}{\pi\xi})}\prod_{p=1}^{\infty}\frac{[R_{p}(i\pi)
R_{p}(o)]^{\frac{1}{2}}}{R_{p}(\beta)},\\
& &\langle 0|\phi^{'}(\beta _{1})\phi^{'}(\beta _{2})|0\rangle =
-lng^{'}(\beta _{2}-\beta _{1}),\ \  
s^{'}(\beta)=\frac{g^{'}
(-\beta)}{g^{'}(\beta)},\ \ g^{'}(\beta)=\frac{\Gamma(1+\frac{i\beta}
{\pi(1+\xi)})}{\Gamma(\frac{1}{1+\xi}+\frac{i\beta}{\pi(1+\xi)})}
\prod_{p=1}^{\infty}\frac{R^{'}_{p}(\beta)}{[R^{'}_{p}(i\pi)R^{'}_{p}(0)]
^{\frac{1}{2}}},\\
& & \langle 0|\phi(\beta _{1})\phi^{'}(\beta _{2})|0\rangle =-lnh(\beta_{2}-
\beta_{1}),\ \ tg(\frac{\pi}{4}+i\frac{\beta}{2})=\frac{h
(\beta)}{h(-\beta)},\ \ h(\beta)=\frac{\Gamma(\frac{1}{4}+\frac{i\beta}{2\pi}
)}{\Gamma(\frac{3}{4}+\frac{i\beta}{2\pi})}\ \ \ .
\end{eqnarray*}
In Eq.(7) the screening operators $\chi ,\chi ^{'}$ are defined through 
\begin{eqnarray*}
& &\langle u|\chi |v\rangle \equiv \eta \langle u|\int_{C}\frac{d\gamma}{2\pi}
\overline{V}(\gamma)|v\rangle,\ \ \ \ \ 
\overline{V}(\gamma)\equiv :e^{-i\overline{\phi}(\gamma)}:\equiv :e^{-i(\phi(
\gamma+\frac{i\pi}{2})+\phi(\gamma -\frac{i\pi}{2}))}:\ \ \ ,
\end{eqnarray*}
$$
\langle u|\chi ^{'}|v\rangle \equiv 
\eta ^{'}\langle u|\int_{C^{'}}\frac{d\gamma}{2\pi}\overline{V}^{'}(\gamma)|v
\rangle ,\ \ \ \ \ 
\overline{V}^{'}(\gamma)\equiv :e^{-i\overline{\phi}^{'}(\gamma)}:\equiv :
e^{-i(\phi^{'}(\gamma+\frac{i\pi}{2})+\phi^{'}(\gamma -\frac{i\pi}{2}))}:\ \ \ .\eqno(7a)
$$

In Eq.(7a) $\langle u|$ and $|v\rangle $ are some states in the Fock space 
and its dual space of the bosonic fields $\phi (\beta)$ and $\phi^{'}(\beta) \ \ ,
 \ \ \eta \ \ ,
\eta ^{'}$ are irrelavent constants$^{[12]}$.The integration contour $C$ $(C^{'})$ is 
taken such that it encloses only the poles originated from the action of 
$\overline{V} \ \ ,(\overline{V}^{'})$ on the right-hand state $|v\rangle $ clockwise
 $^{[12]}$.

In addition to the exchange relation of (5) and (6),the operator product of 
particle-creation operators $Z_{a}(\beta )$ has the 
following singular properties  $^{[12]}$ when $\beta _{2} $ approaches 
$\beta _{1} $ :
\begin{eqnarray}
iZ_{a}(\beta_{2})Z_{b}(\beta_{1})=\frac{C_{a,b}}{\beta_{2}-\beta_{1}-i\pi}+...
\ \ \ \ , \ \ \ \ \beta_{2}\longrightarrow \beta_{1} \ \ \ .
\end{eqnarray}
From the bosonic representation (7) and its regularized version (see sect.4),
Lukyanov got the form-factors of SG model in the bulk case.

\section {Boundary operators in the bosonic version}
Now let us consider SG field theory with an integrable boundary at x=0,
which is described by the action $^{[2,4]}$
\begin{eqnarray}
{\cal L}=\int_{0}^{\infty}dx\int_{-\infty}^{\infty}dt\frac{1}{2}
(\partial_{\mu} \varphi )^{2}-\frac{m^{2}}{\beta ^{2}}cos\beta \varphi 
-\int_{-\infty}^{\infty}dtMcos(\frac{\beta}{2}(\varphi |_{x=0}-\varphi_{0}))
\ \ \ .
\end{eqnarray}
It has been proved that the boundary SG model (9) is integrable$^{[2,3,4]}$ 
and the asymptotic particle states are generated by the same 
set of creation operators $Z_{a}(\beta)$ as in the bulk case in the 
following  way $^{[2]}$:
\begin{eqnarray}
|Z_{a_{1}}(\beta _{1})Z_{a_{2}}(\beta _{2}).... Z_{a_{N}}(\beta _{N})
\rangle _{B}=Z_{a_{1}}(\beta _{1}).... Z_{a_{N}}(\beta _{N})|B\rangle \ \ .
\end{eqnarray} 
But now the vacaum state $|0\rangle$ is replaced by the $|B\rangle $ ,
which is the ground state of SG model with a boundary.The state $|B\rangle$ 
satisfies the relations
\begin{eqnarray}
Z_{i}(\beta)|B\rangle =R^{j}_{i}(\beta)Z_{j}(-\beta)|B\rangle \ \ \ . 
\end{eqnarray}
If such $|B\rangle$ exists,the consistency of (5) and (11) leads to 
the boundary Yang-Baxter equation and the unitary relation 
\begin{eqnarray}
& &R_{2}(\beta _{2})S_{12}(\beta _{1}+\beta_{2})R_{1}(\beta_{1})S_{21}(\beta_{1}-
\beta_{2})=S_{12}(\beta_{1}-\beta_{2})R_{1}(\beta_{1})S_{21}(\beta_{1}+\beta_
{2})R_{2}(\beta_{2}) \nonumber\\
& & R(\beta)R(-\beta)=1 
\end{eqnarray}
which in turn guarantees the integrability of the model$^{[2,3,4,29,30,31]}$.
Supplied (12) 
with the boundary crossing relation given by Ghoshal and Zamolodchikov$^{[2]}$
\begin{eqnarray}
C^{ac}R^{b}_{c}(\frac{i\pi}{2}-\beta)=\sum_{c,d,e}S^{ab}_{cd}(2\beta)C^{de}R^{c}
_{e}(\frac{i\pi}{2}+\beta) \ \ \ \ \ , \ \ \ C=\left( \begin{array}{ll}
0&1\\1&0 \end{array}
\right)  
\end{eqnarray}
one can determine the exact boundary reflecting matrix $R(\beta)$ upon to some ``CDD" 
ambiguity.This ambiguity can be cancelled by some dynamical requirement$^{[1,2]}$.
Thus the crucial problem is to find the ground state $|B\rangle$ .For this 
purpose we first obtain a R-matrix satisfying all the requirements (12) and 
(13). 

In this paper,we only  consider the diagonal solution to (12) and (13),which 
corresponds to the fixed boundary condition $\varphi |_{x=0}=\varphi _{0}$ (or 
 to the case of $M\longrightarrow \infty $ in the action (9) )$^{[2,4]}$.
In this special 
case,minimal solution $R(\beta)$ (which has the minimum poles in the 
physical strip ) reads
\begin{eqnarray}
& &R(\beta)=r(\beta)\left( \begin{array}{ll}1&0\\0&\frac{sh\frac{\mu -\beta}{\xi}}
{sh\frac{\mu+\beta}{\xi}} 
\end{array}
\right) \ \ \ ,\ \ \ r(\beta)=r_{0}(\beta)r_{1}(\beta) \ \ ,\\
& &r_{1}(\beta)=\prod_{p=0}^{\infty}\frac{\Gamma(\frac{\beta -\mu}
{i\pi\xi}+\frac{2p}{\xi})\Gamma(\frac{\beta +\mu}
{i\pi\xi}+\frac{2p+2}{\xi}+1) }
{\Gamma(\frac{-\beta -\mu}{i\pi\xi}+\frac{2p}{\xi})\Gamma(\frac{-\beta +\mu}
{i\pi\xi}+\frac{2p+2}{\xi}+1) }
 \frac{\Gamma(\frac{\mu-\beta}{i\pi\xi}+\frac{2p+1}{\xi}+
1)\Gamma(-\frac{\mu+\beta}{i\pi\xi}+\frac{2p+1}{\xi})}
{\Gamma(\frac{\mu+\beta}{i\pi\xi}+\frac{2p+1}{\xi}+
1)\Gamma(-\frac{\mu-\beta}{i\pi\xi}+\frac{2p+1}{\xi})}\ \ ,
\nonumber\\
& &r_{0}(\beta)=\frac{\Gamma(1+\frac{2i\beta}{\pi\xi})\Gamma(\frac{1}{\xi}-
\frac{2i\beta}{\pi\xi})}{\Gamma(\frac{1}{\xi}+\frac{2i\beta}{\pi\xi})
\Gamma(1-\frac{2i\beta}{\pi\xi)})}
\prod_{p=1}^{\infty}\frac{\Gamma(\frac{4p}{\xi}+\frac{2
i\beta}{\pi\xi})\Gamma(1+\frac{4p}{\xi}+\frac{2i\beta}{\pi\xi})
\Gamma(\frac{4p+1}{\xi}-\frac{2i\beta}{\pi\xi})\Gamma(1+\frac{4p-1}{\xi}
-\frac{2i\beta}{\pi\xi})} {\Gamma(\frac{4p}{\xi}-\frac{2i\beta}
{\pi\xi})\Gamma(1+\frac{4p}{\xi}-\frac{2i\beta}{\pi\xi})
\Gamma(\frac{4p+1}{\xi}+\frac{2i\beta}{\pi\xi})\Gamma(1+\frac{4p-1}{\xi}+
\frac{2i\beta}{\pi\xi})}\ \ .\nonumber
\end{eqnarray}
We find that this solution is the same as the diagonial solution (k=0) 
of Ghoshal and Zamolodchikov$^{[2]}$ if one use the folllowing transformation :
\begin{eqnarray*}
\frac{1}{\xi } \longrightarrow \lambda \ \ \ , -i\beta \longrightarrow \mu 
\ \ \ ,\mu \longrightarrow -i\pi (\xi +\frac{1}{2\lambda })\ \ \ .
\end{eqnarray*}
The diagonal Solution (14) has one formal parameter $\mu$ which can be 
related to the physical parameter $\phi _{0} $ in the 
Lagrangian $^{[2,37]}$.

We then assume that the boundary ground state $|B\rangle$ and its dual state 
$\langle B|$ can be expressed by 
$$
|B\rangle =e^{\Psi_{-}}|0\rangle \ \ \ ,\ \ \ 
\langle B|=\langle 0|e^{\Psi_{+}} \eqno(14a)
$$
where $e^{\Psi_{-}}$ is called the boundary operator and $e^{\Psi_{+}}$ is 
the dual operator of $e^{\Psi_{-}}$ .Similar to the XXZ model with a 
reflection boundary$^{[25]}$,we need further  
\begin{eqnarray}
\langle B|Z^{*}_{i}(-\beta)=\langle B|Z^{*}_{j}(\beta)R^{j}_{i}
(\beta),{\rm where} \ \ Z^{*}_{i}(\beta)=C^{a}_{i}Z_{a}(i\pi+\beta), \ \  C=\left(
\begin{array}{ll}
0&1\\1&0
\end{array}
\right) \ \ .
\end{eqnarray}                                                 

In order to study the properties of the operators $\Psi_{\pm}$,we first 
introduce the following decomposition of the bosonic fields $\phi(\beta)$ and
 $\phi^{'}(\beta)$
\begin{eqnarray*}
\phi(\beta)=\phi_{+}(\beta)+\phi_{0}(\beta)+\phi_{-}(\beta) ,\ \ \phi^{'}
(\beta)=\phi_{+}^{'}(\beta)+\phi_{0}^{'}(\beta)+\phi_{-}^{'}(\beta)\ \ \ ,
\end{eqnarray*}
where $\phi_{\pm}(\beta)$ , $\phi_{\pm}^{'}(\beta)$ , $\phi_{0}$ and 
$\phi_{0}^{'}$ satisfy the following relations$^{[12]}$
\begin{eqnarray*}
& &\phi_{+}(\beta)|0\rangle=\langle 0|\phi_{-}(\beta)=0\ \ , \ \ [\phi_{0},
\phi_{0} ]=0 \ \ ,\ \ [\phi_{0},\phi_{\pm}(\beta)]=0\ \ ,\ \  
[\phi_{\pm}(\beta_{1}),\phi_{\pm}(\beta_{2})]=0 ,\\
& &[\phi_{+}(\beta_{1}),\phi_{-}(\beta_{2})]=-lng(\beta_{2}-\beta_{1})\ \ ,\ \ 
 \phi_{+}^{'}(\beta)|0\rangle =\langle 0|\phi_{-}^{'}(\beta)=0 \ \ ,
\ \ [\phi_{0}^{'},\phi_{\pm}^{'}(\beta)]=0,\\
& &[\phi_{0}^{'},\phi_{0}^{'} ]=0 \ \ ,\ \ 
[\phi_{\pm}^{'}(\beta_{1}),\phi_{\pm}^{'}(\beta_{2})]=0 \ \ \ , 
\ \ [\phi_{+}^{'}(\beta_{1}),\phi_{-}^{'}(\beta_{2})]=-lng^{'}(\beta_{2}-
\beta_{1})\ \ .
\end{eqnarray*}
Actually, $\phi _{\pm}(\beta)$ ($\phi _{\pm}^{'}(\beta))$ are proper limit of 
regularized bosonic field $\phi_{\epsilon\pm}$ ($\phi_{\epsilon\pm}^{'}$
 respectivly ), and $\phi_{0} $ ($\phi_{0}^{'}$) is the limit of zero mode of
 regularized one.The regularized bosonic fields will be discussed in 
the section 4.

Now let us make the ansatz for $\Psi_{\pm}$ :
\begin{eqnarray}
& &[\Psi_{\pm},\phi_{\pm}(\beta)]=0,\ \ [\Psi_{\pm},\phi_{0}]=0,\ \ 
[\Psi_{+},\phi_{-}(i\pi +\beta)]=\kappa_{+}\phi_{+}(i\pi -\beta)-\frac{i}{2}
\gamma_{+}(-\beta) ,\nonumber\\
& &[\Psi_{-},\phi_{+}(\beta)]=\kappa_{-}\phi_{-}(-\beta)-\frac{i}{2}
\gamma_{-}(-\beta) ,\nonumber\\
& &[\kappa_{\pm},everthing]=0 \ \ , \ \ [\gamma_{\pm}(\beta),everything]=0
 \ \ \ .
\end{eqnarray}
Substituting the above ansatz into the  equations (11) and (15) for i=+ in 
(11) and i=- in (15) ),
we find that $\kappa_{\pm}$ and $\gamma_{\pm}$ can be determined by 
\begin{eqnarray}
& &e^{i\phi_{+}(\beta)}|B\rangle =f(\beta)e^{\frac{\beta}{2\xi}}e^{i\phi_{-}
(-\beta)}|B\rangle ,\ \ \langle B|e^{i\phi_{-}(i\pi+\beta)}=e^{\frac{\beta}
{2\xi}}f(-\beta)
sh\frac{\mu+\beta}{\xi}\langle B|e^{i\phi_{+}(i\pi-\beta)},\\
& & {\rm where }\ \ \  r(\beta)=\frac{f(\beta)}{f(-\beta)}\ \ \ \ ,\nonumber\\
& &f(\beta)=\frac{\Gamma(\frac{1}{\xi}-\frac{2i\beta}{\pi\xi})}{\Gamma(1-
\frac{2i\beta}{\pi\xi})}
\prod_{p=1}^{\infty}\frac{\Gamma(\frac{4p+1}{\xi}-\frac{2i
\beta}{\pi\xi})\Gamma(1+\frac{4p-1}{\xi}-\frac{2i\beta}{\pi\xi})}
 {\Gamma(\frac{4p}{\xi}-\frac{2i\beta}{\pi\xi})\Gamma(1+\frac{4p}{\xi}-
\frac{2i\beta}{\pi\xi})}\prod_{p=0}^{\infty}\frac{\Gamma(\frac{\beta-\mu}
{i\pi\xi}+\frac{2p}{\xi})
\Gamma(\frac{\beta+\mu}{i\pi\xi}+\frac{2p+2}{\xi}+1)}{\Gamma(\frac{\beta+\mu}
{i\pi\xi}+\frac{2p+1}{\xi}+1)\Gamma(\frac{\beta-\mu}{i\pi\xi}+
\frac{2p+1}{\xi})} .\nonumber
\end{eqnarray}
From directly calculating ,we  get the 
exact $\kappa_{\pm} $ and $ \gamma_{\pm}$ 
\begin{eqnarray*}
\kappa_{+}=1 \ \ ,\ \ \gamma_{+}(\beta )=-\frac{\beta}{\xi}+2lnf(\beta)
+2lnsh\frac{\mu -\beta }{\xi}-lng(-2\beta) \ \ ,
\end{eqnarray*}
$$
\kappa_{-}=1 \ \ ,\ \ \gamma_{-}(\beta )=-\frac{\beta}{\xi}+2lnf(-\beta)
-lng(2\beta) \ \ \eqno(17a) 
$$

From direct caculating and using the definition of the $\overline{\phi}(\beta)$ 
\begin{eqnarray*}
& &\overline{\phi}(\beta)=\phi (\beta +\frac{i\pi}{2})+
\phi (\beta -\frac{i\pi}{2})\ \ \ ,
\end{eqnarray*}
one can get 
\begin{eqnarray}
& &e^{-i\overline{\phi}_{+}(\beta)}|B\rangle=\frac{g(-2\beta-i\pi)g(-2\beta)
g(-2\beta+i\pi)}{f(\beta-i\frac{\pi}{2})f(\beta+i\frac{\pi}{2})}e^{-\frac{
\beta}{\xi}}e^{-i\overline{\phi}_{-}(-\beta)}|B\rangle  \ \ ,\\
& &\langle B|e^{-i\overline{\phi}_{-}(i\pi+\beta)}=\frac{g(2\beta+i\pi)
g(2\beta)g(2\beta-i\pi)}{f(-\beta-i\frac{\pi}{2})f(-\beta+i\frac{\pi}{2})
sh\frac{\mu+\beta+i\frac{\pi}{2}}{\xi}sh\frac{\mu+\beta-i\frac{\pi}{2}}{\xi}}
\langle B|e^{-i\overline{\phi}_{+}(i\pi-\beta)} \ \ .\nonumber
\end{eqnarray}
Then we must check the consistence of our ansatz for the second equation of 
(11) and (15) i.e. i=- for (11) and i=+ for (15) ).Using the explicit form of
 $Z_{-}(\beta)$ in (7) and (17),(18), 
we find the second equation of (11) is equivalent to the following equation
\begin{eqnarray}
& &\int_{C}d\eta \frac{\Gamma(\frac{\mu+\eta}{i\pi\xi}+\frac{1}{2\xi}+1)
\Gamma(-\frac{1}{2\xi}-i\frac{\beta-\eta}{\pi\xi})
\Gamma(-\frac{1}{2\xi}+i\frac{\beta-\eta}{\pi\xi})\Gamma
(-\frac{1}{2\xi}-i\frac{\beta+\eta}{\pi\xi})} {\Gamma(\frac{\eta-\mu}{i\pi\xi}-
\frac{1}{2\xi})\Gamma(1+\frac{1}{2\xi}-i\frac{\beta+\eta}{\pi\xi})}
 \nonumber \\
& &sh\frac{\mu+\beta}{\xi}\eta(\eta^{2}+\frac{\pi^{2}}{4})
:e^{-i\overline{\phi}_{-}(\eta)-i\overline{\phi}_{-}(-\eta)}:|B\rangle 
=(\beta \longleftrightarrow -\beta)
\end{eqnarray}
where the integration contour C encloses the poles $-i\pi\xi(n+1)-i\frac{\pi}
{2}-\mu $ , $ -i\pi\xi n+i\frac{\pi}{2}+\beta$ ,$ -i\pi\xi n+i
\frac{\pi}{2}-\beta $ but not the poles $i\pi \xi n-i\frac{\pi}{2}+\beta $
$ ( n \geq 0 ) $ .We change the integration parameter $\eta$ to $-\eta$ ,and 
with a corresponding new contour $C^{'}$,we still get the same integral as 
the LHS of (19).We find that the corresponding new integration
contour $C^{'}$ can be deformed to the same as C. Adding this integral to 
LHS of (19) we have 
\begin{eqnarray*}
2\times LHS  \ \  of \ \ (19)
&=&\int_{C}d\eta \frac{ sh\frac{\mu-\beta}{\xi}sh\frac{\mu 
+\beta}{\xi}\pi \eta(\eta^{2}+\frac{\pi^{2}}{4})sh\frac{2\eta}{\xi}
}
{sin(-\frac{\pi}{2\xi}+i\frac{\mu-\eta}{\xi})sin(-\frac{\pi}
{2\xi}+i\frac{\mu-\eta}{\xi})}\\
& &\frac{\Gamma(-\frac{1}{2\xi}-i\frac{\beta-\eta}{\pi\xi})\Gamma(-\frac{1}{2\xi}+i
\frac{\beta-\eta}{\pi\xi})\Gamma(-\frac{1}{2\xi}-i\frac{\beta+\eta}{\pi\xi}
)\Gamma(-\frac{1}{2\xi}+i\frac{\beta+\eta}{\pi\xi}):e^{-i\overline{\phi}_{-}
(\eta)-i\overline{\phi}_{-}(-\eta)}:|B\rangle }{\Gamma(i\frac{\mu-\eta}{\pi\xi}-\frac{1}{2\xi})\Gamma(i\frac{\mu+\eta}{\pi\xi}
-\frac{1}{2\xi})}
\end{eqnarray*}
$$
\ \ \ \ \eqno(19a)
$$
which is symmetric under $\beta \longleftrightarrow -\beta $ .Using the same 
method,it is easy to check that the second equation of (15) also has the 
same symmetry under $\beta \longleftrightarrow -\beta$ .Thus,the boundary 
operator of fixed case is obtained,if (16) can be satisfied for some 
$\Psi_{\pm}$ .It will be shown that such $\Psi_{\pm}$  can be 
obtained in section 4 when one takes the limit of regularized 
operators $\Psi_{\epsilon\pm}$ as $\epsilon \longrightarrow 0$.

To study the boundary bound states introduced by Ghoshal and 
Zamolodchikov,
we now analyze the physical poles (in the 
physical strip $0<Im(\beta)<\pi) $ of the reflecting matrix $R(\beta)$ 
.Poles of $r_{0}(\beta )$ term (in eq.(14) ) are associated with 
breathers and do not correspond to the boundary (new) bound states. The other 
poles of $r_{1}(\beta )$ term correspond to the boundary bound states.
For $-\frac{\pi\xi}{2}\leq Im(\mu) \leq0$
( which is corresponding to Ghoshal and Zamolodchikov's $0\leq \xi \leq 
\frac{\pi}{2}$ in Ref.[2]),there is no pole in the physical strip.
For $0<Im(\mu)\leq [\frac{1}{\xi}]\pi\xi $ ,two series of poles appear in the 
physical strip : $\mu -i\pi \xi l \ \ $ and $i\pi -\mu +i\pi \xi l \ \ \ (0 \leq l \leq \lambda
 \ \ , \ \ \lambda=[\frac{Im(\mu)}{\pi\xi}] )$ .The structure of physical 
poles is the same as the results of Skorik and 
Saleur et al in Ref.[4],where some semi-classical explanation of boundary 
bound states was given. The $i\pi-\mu +i\pi\xi l $ 
 series correspond to the ``cross-channel" poles of kink and 
antikink reflecting at the boundary.Hence we need only to study the bound 
state corresponding to the poles $\mu-i\pi\xi l\ \ $ 
and denote them by $\ \ |B\rangle ^{(+l)} $ .
Using the boundary bootstrap methods$^{[2]}$,we can obtain the corresponding 
boundary reflecting matrices $R^{(l)}(\beta)$,which describe the reflection 
between kink or antikink and the boundary bound states $\ \ |B\rangle ^{(+l)}
 $ :
\begin{eqnarray*}
& &Z_{i}(\beta )|B\rangle ^{(+l)}=R^{(l)j}_{ \ \ \ i}(\beta )|B\rangle ^{(+l)}\ \ \ \ ,\\
& &R^{(l)}(\beta)=r(\beta)s(\beta+\mu-i\pi\xi l)s(\beta-\mu +i\pi\xi l))
\left( \begin{array}{lc}1&0\\0&\frac{sh\frac{\mu-i\pi-\beta}{\xi}}{sh\frac{
\mu-i\pi +\beta}{\xi}}
\end{array}
\right) =r^{(l)}(\beta)\left( \begin{array}{lc}1&0\\0&\frac{sh\frac{\mu-i\pi-
\beta}{\xi}}{sh\frac{\mu-i\pi +\beta}{\xi}}
\end{array}
\right),
\end{eqnarray*}
where $0\leq l \leq \lambda $ .One can check that the reflecting matrices $
R^{(l)}(\beta)$ satisfy the boundary unitary and crossing condition (12) and 
(13), which are equivalent to the equations 
\begin{eqnarray*}
r^{(l)}(\beta)r^{(l)}(-\beta)=1\ \ ,\ \ \ r^{(l)}(\frac{i\pi}{2}-\beta)=s(2\beta)\frac{sh\frac{i\pi+2\beta}{\xi}sh\frac{
\mu-i\pi-\beta+\frac{i\pi}{2}}{\xi}}{sh\frac{i\pi-2\beta}{\xi}sh\frac{\mu
-i\pi+\beta+\frac{i\pi}{2}}{\xi}}r^{(l)}(\frac{i\pi}{2}+\beta) \ \ .
\end{eqnarray*}
With these $R^{(l)}(\beta)$ , one can construct the corresponding boundary 
bound  states $|B\rangle^{(+l)}$ and $\langle B|^{(+l)}$ by the 
same method as that we already use to construct $|B\rangle$  and $\langle B|$ 
from $R(\beta)$.
On the other hand,due to $|B\rangle^{(+l)}$ 
$ \ \ (\langle B|^{(+l)})$ being the bound states of $|B\rangle \ \ \ 
(\langle B|$) with  an incoming particle ( kink ) carrying rapidity 
$\mu -i\pi\xi l$ ,they can be obtained through the 
following way
\begin{eqnarray}
|B\rangle^{(+l)} =Res_{\beta=\mu-i\pi\xi l}Z_{+}(\beta)|B\rangle \ \ , \ \ 
\langle B|^{(+l)} =Res_{\beta=\mu-i\pi\xi l}\langle B|Z^{*}_{+}(-\beta) \ \ .
\end{eqnarray}
The boundary bound state has been discussed also
by Saleur and Skorik et al in Ref.[4] and Ref.[40].

The $R(\beta)$ and $R^{(l)}(\beta)$ $(0\leq l \leq \lambda)$ describe the 
reflection of kink and antikink with the corresponding boundary.However, 
there exist
other particles --- breathers $B_{n}$ $(0<n\leq [\frac{1}{\xi}])$ in SG
 model.The breathers are considered as the bound states of kink and antikink 
pair and are defined as following
\begin{eqnarray*}
B_{n}(\beta)=Z_{+}(\beta-\frac{i\pi}{2}+\frac{i\pi\xi n}{2})Z_{-}(\beta
+\frac{i\pi}{2}-\frac{i\pi\xi n}{2})+(-1)^{n}Z_{-}(\beta-\frac{i\pi}{2}
+\frac{i\pi\xi n}{2})Z_{+}(\beta+\frac{i\pi}{2}-\frac{i\pi\xi n}{2})
\end{eqnarray*}
where $0<n\leq [\frac{1}{\xi}]$ .
Using these definitions ,one can get the reflecting matrices of breathers
 with the boundary and its bound states
\begin{eqnarray*}
& &B_{n}(\beta)|B\rangle=s(2\beta)r(\beta+\frac{i\pi}{2}-\frac{i\pi\xi n
}{2})r(\beta-\frac{i\pi}{2}+\frac{i\pi\xi n}{2})
\frac{sh\frac{i\pi+2\beta}
{\xi}sh\frac{\mu -\beta +\frac{i\pi}{2}-\frac{i\pi\xi n}{2}}{\xi}}
{sh\frac{i\pi-2\beta}{\xi}sh\frac{\mu +\beta +\frac{i\pi}{2}-\frac{i\pi\xi n}
{2}}{\xi}}B_{n}(-\beta)|B\rangle\\
& &B_{n}(\beta)|B\rangle^{(+l)}=s(2\beta)r^{(l)}(\beta+\frac{i\pi}{2}-
\frac{i\pi\xi n}{2})r^{(l)}(\beta-\frac{i\pi}{2}+\frac{i\pi\xi n}{2})
\frac{sh\frac{i\pi+2\beta}
{\xi}sh\frac{\mu -\beta -\frac{i\pi}{2}-\frac{i\pi\xi n}{2}}{\xi}}
{sh\frac{i\pi-2\beta}{\xi}sh\frac{\mu +\beta -\frac{i\pi}{2}-\frac{i\pi\xi n}
{2}}{\xi}}B_{n}(-\beta)|B\rangle^{(+l)} \ \ .
\end{eqnarray*}

\section{Boundary operators versus q-oscillator}
In sect.3,we give the relation of the boundary operators $\Psi _{\pm }$ with 
the bosonic fields $\phi_{\pm}(\beta)$ in (16).If such $\Psi_{\pm}$ exist,we 
can construct a boundary state $|B\rangle$ and its dual $\langle B|$ via 
(14a) which ensures the boundary relations (11) and (15).However,it is 
still unknown how the boundary operators  $\Psi _{\pm }$ can be explicitely 
expressed in the term of bosonic fields $\phi (\beta )$ and $\phi '(\beta )$ .
Fortunately,in order to calculate form-factors 
in the bulk case ,Lukyanov 
introduced some regularized field  $\phi _{\epsilon }(\beta )$ and $ 
\phi' _{\epsilon }(\beta )$ ,which can be expanded into an infinite sum of 
q-oscillator mode $^{[12]}$.In this papper,with the helps of q-oscillator 
expansion of $\phi _{\epsilon }(\beta )$ ,
we can obtain the 
explicite form of regularized boundary operators $\Psi _{\pm \epsilon }$ 
in terms of the q-oscillators.This not only proves the existence of 
$\Psi_{\pm}$ but also permits one to calculate the form-factors  in the 
regularized version.After $\epsilon \longrightarrow 0 $ ,the form-factors of 
SG model with fixed boundary are obtained (see sect.5).In the following of 
this section ,we will give boundary operators in term of  q-oscillator
(The regularized bosonic field itself has a very important meaning---it is related 
to the bosonization of XYZ model$^{[14,26]}$,and has some relation with the 
deformed Virasoro algebra$^{[14,32,33,34]}$ ).

The bosonic fields $\phi(\beta)$ and $\phi^{'}(\beta)$ in sect.2  cannot be 
expanded into an infinite sum of oscillator modes as the usual free fields
does.However, we can perform an oscillator realization by the ultraviolate
regularization introduced by Lukyanov $^{[12]}$ .To do this,the first 
thing is to introduce a regularization parameter $\epsilon$ such
 that $-\frac{\pi}{\epsilon}\leq \beta \leq \frac{\pi}{\epsilon}$ and then 
consider regularized fields $\phi_{\epsilon}(\beta)$ and $\phi_{\epsilon}
^{'}(\beta)$ which can be expanded by the q-oscillators.Then the bosonic fields $\phi(\beta)  \ \ \phi^{'}(\beta)$ can be 
viewed as proper limit of $\phi_{\epsilon}(\beta) \ \ \phi^{'}_{\epsilon}(\beta)
$ as $\epsilon \longrightarrow 0$ .The oscillator expansion of the regularized
 bosonic fields are $^{[12]}$
\begin{eqnarray}
& &\phi_{\epsilon}(\beta)=\sqrt{\frac{1+\xi}{2\xi}}(Q-\epsilon \beta P)+\sum_{m
\not=0}\frac{a_{m}}{ish\pi m \epsilon}exp(im\epsilon\beta),\nonumber \\
& &\phi_{\epsilon}^{'}(\beta)=-\sqrt{\frac{\xi}{2(1+\xi}}(Q-\epsilon \beta P)-
\sum_{m\not=0}\frac{sh\frac{\pi m\epsilon\xi}{2}}{ish\pi m \epsilon 
sh\frac{\pi m\epsilon (1+\xi)}{2}}a_{m}exp(im\epsilon\beta),\nonumber \\
& &\overline{\phi}_{\epsilon}(\beta)=\phi_{\epsilon}(\beta +\frac{i\pi}{2})
\phi_{\epsilon}(\beta -\frac{i\pi}{2})\ \ ,\ \ 
\overline{\phi}_{\epsilon}^{'}(\beta)=\phi_{\epsilon}^{'}(\beta +\frac{i\pi}{2})
\phi_{\epsilon}^{'}(\beta -\frac{i\pi}{2})\ \ ,\nonumber\\
& &[P,Q]=-i, \ \ [P,a_{m}]=0, \ \ [Q,a_{m}]=0,\ \ [a_{m},a_{n}]
=sh\frac{m\pi\epsilon}{2}shm\pi\epsilon\frac{sh\frac{m\pi\epsilon
(1+\xi)}{2}}{msh\frac{m\pi\epsilon\xi}{2}}\delta_{m+n,0},\nonumber \\
& &V_{\epsilon}(\beta)=:e^{i\phi_{\epsilon}(\beta)}: \ \ , \ \ V_{\epsilon}^{'}
(\beta)=:e^{i\phi_{\epsilon}^{'}(\beta)}: \ \ , \ \ \overline{V}_
{\epsilon}(\beta)=:e^{-i\phi_{\epsilon}(\beta +\frac{i\pi}{2})-i\phi_{\epsilon}
(\beta-\frac{i\pi}{2})}:\ \ ,\nonumber \\
& &\overline{V}_{\epsilon}^{'}(\beta)=:e^{-i\phi_{\epsilon}^{'}(\beta +\frac{i
\pi}{2})-i\phi_{\epsilon}^{'}(\beta-\frac{i\pi}{2})}: \ \ \ , \ \ \ \ 
 \stackrel{\sim }{V}_{\epsilon}^{'}(\beta)=:e^{i\phi_{\epsilon}^{'}(\beta +
\frac{i\pi}{2})+i\phi_{\epsilon}^{'}(\beta-\frac{i\pi}{2})}: \ \ \ \ .
\end{eqnarray}
Similarly,we can introduce a regularized boundary reflecting matrix is 
$R_{\epsilon}(\beta)$ 
\begin{eqnarray*}
& &R_{\epsilon}(\beta)=\frac{f_{\epsilon}(\beta)}{f_{\epsilon}(-\beta)}
\left( \begin{array}{ll}1&\\
&\frac{\Gamma_{\frac{1}{2}\epsilon\xi}(1-
\frac{\mu+\beta}{i\pi\xi})\Gamma_{\frac{1}{2}\epsilon\xi}(\frac{\mu+\beta}{
i\pi\xi})}{\Gamma_{\frac{1}{2}\epsilon\xi}(1-\frac{\mu-\beta}{i\pi\xi})
\Gamma_{\frac{1}{2}\epsilon\xi}(\frac{\mu-\beta}{i\pi\xi})}
\end{array}
\right)\ \ \ ,\\
& &f_{\epsilon}(\beta)=\frac{(e^{-4\pi\epsilon +2i\epsilon \beta};e^{-\pi
\epsilon\xi},e^{-4\pi\epsilon})(e^{-\pi\epsilon\xi -2i\epsilon \beta};
e^{-\pi\epsilon\xi},e^{-4\pi\epsilon})}{(e^{-\pi\epsilon +2i\epsilon \beta};
e^{-\pi\epsilon\xi},e^{-4\pi\epsilon})(e^{-\pi\epsilon\xi -3\pi\epsilon+
2i\epsilon \beta};e^{-\pi\epsilon\xi},e^{-4\pi\epsilon})}\\
& & \ \ \ \times \frac{(e^{-\pi\epsilon+i\epsilon(\beta-\mu)};e^{-\pi\epsilon
\xi},e^{-2\pi\epsilon})(e^{-\pi\epsilon\xi-\pi\epsilon+i\epsilon(\beta+\mu)};
e^{-\pi\epsilon\xi},e^{-2\pi\epsilon})} {(e^{i\epsilon(\beta-\mu)};e^{-\pi
\epsilon\xi},e^{-2\pi\epsilon})(e^{-\pi\epsilon\xi-2\pi\epsilon+i\epsilon(
\beta+\mu)};e^{-\pi\epsilon\xi},e^{-2\pi\epsilon})} \ \ \ \ \ ,\\
& &\frac{\Gamma_{\frac{1}{2}\epsilon\xi}(1-
\frac{\mu+\beta}{i\pi\xi})\Gamma_{\frac{1}{2}\epsilon\xi}(\frac{\mu+\beta}{
i\pi\xi})}{\Gamma_{\frac{1}{2}\epsilon\xi}(1-\frac{\mu-\beta}{i\pi\xi})
\Gamma_{\frac{1}{2}\epsilon\xi}(\frac{\mu-\beta}{i\pi\xi})}
=\frac{(e^{-\pi\epsilon\xi-i\epsilon(\mu-\beta)};e^{-\pi\epsilon\xi})
(e^{i\epsilon(\mu-\beta)};e^{-\pi\epsilon\xi})}{(e^{-\pi\epsilon\xi-i\epsilon(
\mu+\beta)};e^{-\pi\epsilon\xi})(e^{i\epsilon(\mu+\beta)};
e^{-\pi\epsilon\xi})}\ \ \ \ ,\\
& &(z;q)=\prod_{n=0}^{\infty}(1-zq^{n}) \ \ ,\ \ \ \ 
(z;q_{1},q_{2})=\prod_{n_{1},n_{2}=0}^{\infty}(1-zq_{1}^{n_{1}}q_{2}^{n_{2}})\ \ .
\end{eqnarray*}
$$
\ \ \ \ \eqno(21a)
$$
The reflection matrix $R(\beta)$ in (14) can be viewed as the limit of 
the regularized one in (21a).
The operators $ \Psi_{\pm} $ in sect.3 can also be viewed as the limit 
of the regularized operators $\Psi_{\pm \epsilon } $ .
We assume the forms of the q-oscillator expansion of $\Psi_{\epsilon\pm} $ 
are
\begin{eqnarray}
& &\Psi_{\epsilon-}=\sum_{m=1}^{\infty}\frac{msh\frac{\pi m\epsilon\xi}{2} \ \ 
\alpha_{m}}{2sh\frac{\pi m\epsilon}{2}sh\pi m\epsilon sh\frac{\pi m\epsilon
(\xi +1)}{2}}a_{-m}^{2} +\frac{msh\frac{\pi m\epsilon\xi}{2} \ \ 
\lambda_{m}}{2sh\frac{\pi m\epsilon}{2}sh\pi m\epsilon sh\frac{\pi m\epsilon
(\xi +1)}{2}}a_{-m}-\frac{1}{\epsilon\sqrt{2\xi(1+\xi)}}Q \ \ ,\nonumber\\
& &\Psi_{\epsilon+}=\sum_{m=1}^{\infty}\frac{msh\frac{\pi m\epsilon\xi}{2} \ \ 
\sigma_{m}}{2sh\frac{\pi m\epsilon}{2}sh\pi m\epsilon sh\frac{\pi m\epsilon
(\xi +1)}{2}}a_{m}^{2} +\frac{msh\frac{\pi m\epsilon\xi}{2} \ \ 
\rho_{m}}{2sh\frac{\pi m\epsilon}{2}sh\pi m\epsilon sh\frac{\pi m\epsilon
(\xi +1)}{2}}a_{m}+\frac{1}{\epsilon\sqrt{2\xi(1+\xi)}}Q .\nonumber\\
& & 
\end{eqnarray}
(The zero mode Q term of $\Psi_{\epsilon\pm}$ in this expansion is to reduce 
the factor $e^{\frac{\beta}{2\xi}}$ appearing in (17)).The constants 
$\alpha_{m}$ ,$\lambda_{m}$ ,$\sigma_{m}$ ,$\rho_{m}$ are determined by 
relations
\begin{eqnarray}
& &e^{i\phi_{\epsilon +}(\beta)}e^{\Psi_{\epsilon -}}|0\rangle=f_{\epsilon}(-\beta)
e^{i\phi_{\epsilon -}(-\beta)}e^{\Psi_{\epsilon -}}|0\rangle \ \ ,\\
& &\langle0|e^{\Psi_{\epsilon +}}e^{i\phi_{\epsilon-}(i\pi+\beta)}=f_{\epsilon}
(-\beta)\frac{(e^{-\pi\epsilon\xi-i\epsilon\mu-i\epsilon\beta};e^{-\pi\epsilon
\xi})}{(e^{i\epsilon\mu-i\epsilon\beta};e^{-\pi\epsilon\xi})}
\langle0|e^{\Psi_{\epsilon +}}e^{i\phi_{\epsilon+}(i\pi-\beta)}\ \ \ .
\nonumber
\end{eqnarray}

Substituting (22) into (23),we find  
\begin{eqnarray}
& &\lambda_{m}=-\frac{sh\frac{\pi m\epsilon}{2}sh\frac{\pi m\epsilon(-1-\xi+\frac
{2i\mu}{\pi})}{2}}{2msh\frac{\pi m\epsilon\xi}{2}}
+\theta_{m}(\frac{sh\frac{\pi m\epsilon}{2}sh\frac{\pi m\epsilon(1+\xi)}{2}}
{msh\frac{\pi m\epsilon\xi}{2}}-\frac{sh\frac{3\pi m\epsilon}{2}sh\frac{\pi m
\epsilon(1-\xi)}{2}}{2msh\frac{\pi m\epsilon\xi}{2} ch\pi m\epsilon}) 
 \ \ , \nonumber \\
& &\rho_{m}=\{(\frac{sh\frac{\pi m\epsilon}{2}sh\frac{\pi m\epsilon(-1-\xi+\frac
{2i\mu}{\pi})}{2}}{2msh\frac{\pi m\epsilon\xi}{2}}
-\frac{sh\pi m\epsilon sh\frac{\pi m\epsilon(\xi+\frac
{2i\mu}{\pi})}{2}}{msh\frac{\pi m\epsilon\xi}{2}})\nonumber\\
& &\ \ \ -\theta_{m}(\frac{sh\frac{\pi m\epsilon}{2}sh\frac{\pi m\epsilon
(1+\xi)}{2}}
{msh\frac{\pi m\epsilon\xi}{2}}-\frac{sh\frac{3\pi m\epsilon}{2}sh\frac{\pi m
\epsilon(1-\xi)}{2}}{2msh\frac{\pi m\epsilon\xi}{2} ch\pi m\epsilon})\}
e^{-\pi m\epsilon} \ \ ,\nonumber \\
& &\alpha_{m}=-1 \ \ ,\ \ \sigma_{m}=-e^{-2\pi m\epsilon}\ \ ,\ \
 \theta _{m}=\left\{ \begin{array}{ll}
0  \ \ \ \  & m \ \ \ \   odd\\
1 & m \ \ \ \   even
\end{array}
\right.
\ \ \ .
\end{eqnarray}
Similarly,the regularized versions of vertex operators $V_{\epsilon}(\beta )$,
$V'_{\epsilon}(\beta )$,screening operators $\chi_{\epsilon} $,
$\chi'_{\epsilon}$ can be constructed and 
$V(\beta )$ ,$V'(\beta )$ ,$\chi$ ,$\chi'$ are the limit 
of their regularized counterparts when $\epsilon \longrightarrow 0$.
In the regularized version,vertex 
operators, the screening operators and the boundary operators are all 
expanded by the q-oscillators.Replacing $\Psi_{\pm}$ ,$\phi_{\pm}(\beta)$ ,
$\phi_{0}$ ($\phi_{0} \propto Q $ ) by 
$\Psi_{\epsilon\pm}$ ,$\phi_{\epsilon\pm}(\beta)$ ,
$ Q $ in (16),one find that (16) is satisfied after taking the limit 
$\epsilon \longrightarrow 0$ .The $\kappa_{\pm}$ and $\gamma_{\pm}$ are 
the same as (17a) because the limit of (23) is equivelent to (17).
We thus prove the existence of $\Psi_{\pm}$ and complete the construction 
of the boundary states $|B\rangle$ and $\langle B|$ .

\section{Form-factors in the boundary SG model}
Our final goal is to obtain the form-factors of SG model with boundary.In 
principle,these form-factors enable one to calculate all quantities of 
this quantum field theory$^{[5,6,10,12]}$ .A form-factor is a matrix element 
of a local field operator between two asymptotic states
\begin{eqnarray*}
& &f(\beta_{1},....,\beta_{n})=\langle B|OZ(\beta_{1}).....Z(\beta_{n})
|B\rangle \ \ \ \ \ .
\end{eqnarray*}
The problem of description of all the local operators 
can be considered in the framework of Ref.[12] and Ref.[23] 
 .The solution to this problem is `` a local operator is everything commuting 
 with the type II operators (or the asymptotic generators) "$^{[23]}$.
For SG model,Lukyanov $^{[12]}$ already got this solution : he got three 
generator function operators of local operator $\Lambda_{s}(\alpha )$ 
$s=(-1,0,1)$  
\begin{eqnarray*}
& &\Lambda _{1}(\alpha)=\stackrel{\sim}{V}^{'}(\alpha) \ \ ,\ \ 
\Lambda _{0}(\alpha)=-\frac{i}{(2cos\frac{\pi}{\xi+1})^{\frac{1}{2}}}
[q^{'}\chi^{'}\stackrel{\sim}{V}^{'}(\alpha)-q^{'-1}\stackrel{\sim}{V}^{'}
(\alpha)\chi^{'}] \ \ ,\\
& &\Lambda _{-1}(\alpha)=-\frac{1}{(2cos\frac{\pi}{\xi+1})}
[q^{'}\chi^{'2}\stackrel{\sim}{V}^{'}(\alpha)-q^{'-1}\stackrel{\sim}{V}^{'}
(\alpha)\chi^{'2}
-(q^{'}+q^{'-1})\chi^{'}\stackrel{\sim}{V}^{'}(\alpha)\chi^{'}] 
\end{eqnarray*} 
$$
\ \ \ \ .\eqno(25a)
$$
In this equation  $\stackrel {\sim}{V}'(\alpha )$ is the limit of 
$\stackrel {\sim}{V}'_{\epsilon }(\alpha )$ in (21) as $\epsilon 
\longrightarrow 0 $ ,and the screening operator $\chi '$ is defined in 
sect.2. Using the exchange relation (6), one can find that 
$\Lambda_{m}(\beta)$ satisfy the following relations with 
the particle-creation operator $Z_{a}(\beta)$ 
\begin{eqnarray}
\Lambda_{m}(\alpha)Z_{a}(\beta)=(-1)^{m}Z_{a}(\beta)\Lambda_{m}(\alpha)\ \ \ .
\end{eqnarray}
Thus,in SG model with a boundary a form-factor can be expressed as by 
multipoint functions of the generator functoin operators $\Lambda_{s}$ 
(s=-1,0,1)$^{[12]}$ and the particle-creation operators $Z_{a}$ (a=+,-),
between ground state $|B\rangle $ and its dual $\langle B|$ .We can 
alternatively define the form-factors as
\begin{eqnarray}
F^{m_{1},...,m_{k}}_{a_{1},...,a_{n}}
(\alpha_{1},...,\alpha_{k}| \beta_{1},...,\beta_{n})=
\frac{\langle B|\Lambda_{m_{k}}(\alpha_{k}).... \Lambda_{m_{1}}(\alpha_{1})
Z_{a_{n}}(\beta_{n})....Z_{a_{1}}(\beta_{1})|B\rangle }{\langle B|B\rangle }
 \ \ \ .
\end{eqnarray}

Using exchange relations (5) ,(25) ,the reflection properties (11) ,(15) 
and the singular properties 
(8) ,one can find that the form-factors in the boundary case defined above 
satisfy the following difference equations and the residue condition 
(cf Eq.(1) and (2)),
which are quite similar to the equations appearing in the solvable lattice model 
with integrable boundary$^{[26]}$:

{\large \bf 1.Riemann-Hilbert problem.}
\begin{eqnarray}
& &F^{m_{1},...,m_{k}}_{a_{1},..,a_{i},..,a_{j},..,a_{n}}(\alpha_{1},...,
\alpha_{k}|\beta_{1},...,\beta_{i},...,\beta_{j},...,\beta_{n})=\nonumber\\
& &\ \ \ \ \ \ \ \ S^{a'_{i},a'_{j}}_{a_{i},a_{j}}(\beta_{i}-\beta_{j})F^{m_{1},...,m_{k}}
_{a_{1},..,a'_{j},..,a'_{i},..,a_{n}}(\alpha_{1},...,
\alpha_{k}|\beta_{1},...,\beta_{j},...,\beta_{i},...,\beta_{n})\nonumber\\
& &F^{m_{1},...,m_{k}}_{a_{1},...,a_{n}}
(\alpha_{1},...,\alpha_{k}| \beta_{1},...,\beta_{n})=
R^{a'_{1}}_{a_{1}}(\beta_{1})F^{m_{1},...,m_{k}}_{a_{1},...,a_{n}}
(\alpha_{1},...,\alpha_{k}| -\beta_{1},...,\beta_{n})\nonumber\\
& &F^{m_{1},...,m_{k}}_{a_{1},...,a_{n}}
(\alpha_{1},...,\alpha_{k}| \beta_{1},...,\beta_{n})=
C^{a'_{n}}_{j}R^{j}_{l}(i\pi -\beta_{n})C^{l}_{a_{n}}
F^{m_{1},...,m_{k}}_{a_{1},...,a'_{n}}
(\alpha_{1},...,\alpha_{k}| \beta_{1},...,2i\pi -\beta_{n})
\end{eqnarray}
(The first equation is the result of exchange relation (5),the second 
one is the result of reflection property (11), and the last one is 
the result of Eq.(15). )

{\large \bf 2.Residue condition.}
\begin{eqnarray}
& &2i\pi Res_{\beta_{n}=\beta_{n-1}+i\pi}F^{m_{1},...,m_{k}}_{a_{1},...,
a_{n}}(\alpha_{1},...,\alpha_{k}| \beta_{1},...,\beta_{n})=
F^{m_{1},...,m_{k}}_{a_{1},...,
a_{n-2}}(\alpha_{1},...,\alpha_{k}| \beta_{1},...,\beta_{n-2})
C_{a_{n-1},a_{n}}\nonumber\\
& &\ \ -C^{a'_{n}}_{j}R^{j}_{l}(-\beta_{n-1})C^{l}_{a_{n}}S^{b_{n-1},a'_{n-1}}
_{a'_{n},a_{n-1}}(i\pi -2\beta_{n-1})....S^{b_{1},a'_{1}}_{b_{2},a_{1}}(i\pi 
-\beta_{n-1}-\beta_{1})R^{b'_{1}}_{b_{1}}(\beta_{n-1}-i\pi )\nonumber\\
& &\ \ S^{a"_{1},b"_{1}}_{a'_{1},b'_{1}}(i\pi +\beta_{1}-\beta_{n-1})
....S^{a"_{n-2},b"_{n-2}}_{a'_{n-2},b'_{n-2}}(i\pi +
\beta_{n-2}-\beta{n-1})C_{a'_{n-1},b"_{n-2}}\nonumber\\
& &\ \ \ \ F^{m_{1},.....,m_{k}}
_{a"_{1},....,a"_{n-2}}(\alpha_{1},...,\alpha_{k}|
\beta_{1},...,\beta_{n-2})  \ \ \ .
\end{eqnarray}
(It is the result of singular property (8) with (5) ,(11) ,(15) ,(25). )
There are additional poles corresponding to bound particle (breathers) and 
the boundary bound states.For the poles corresponding to breathers, the 
residue relation are the same as in the bulk case$^{[10]}$. However, 
the poles corresponding to boundary bound states is dependent upon the 
boundary condition. For the fixed boundary case 
which we study in this paper, the form-factors 
between the bound state $|B\rangle^{(+l)}$ and its dual states 
$\langle B|^{(+l)}$ can be directly obtained by the residue form 
as follow
\begin{eqnarray}
& &\langle B|^{(+l)}\Lambda_{m_{k}}(\alpha_{k}).... \Lambda_{m_{1}}Z_{a_{n}}
(\beta_{n})....Z_{a_{1}}(\beta_{1})|B\rangle ^{(+l)}\nonumber\\
& &\ \ \ =\frac{Res_{\beta=\mu -i\pi\xi l}\langle B|Z^{*}_{+}(-\beta)
\Lambda_{m_{k}}(\alpha_{k}).... \Lambda_{m_{1}}Z_{a_{n}}
(\beta_{n})....Z_{a_{1}}(\beta_{1})
Z_{+}(\beta)|B\rangle}{Res_{\beta=\mu -i\pi \xi l}
\langle B|Z_{+}^{*}(-\beta)Z_{+}(\beta)|B\rangle} \ \ .
\end{eqnarray} 
The above analogues of difference equations and residue condition 
also appear in other intergrable field theories with integrable 
boundary,e.g., in SU(2)-ITM with an integrable boundary
$^{[36]}$. 

From the equation (27),one can obtain the quantum boundary 
Knizhnik-Zamolodchikov (KZ) equation
\begin{eqnarray}
& &F^{m_{1},...,m_{k}}_{a_{1},...,a_{n}}
(\alpha_{1},...,\alpha_{k}| \beta_{1},...,\beta_{n})=
R^{b_{1}}_{a_{1}}(\beta_{1})S^{a'_{2},b_{2}}_{a_{2},b_{2}}(\beta_{2}
+\beta_{1})...S^{a'_{n},b_{n}}_{a_{n},b_{n-1}}(\beta_{n}+\beta_{1})
\nonumber\\
& &\ \ \ \  C^{b'_{n}}_{j}R^{j}_{l}(i\pi +\beta_{1})C^{l}_{b_{n}}S^{b"_{n-1},
a"_{n}}_{b'_{n},a'_{n}}(2i\pi +\beta_{1}-\beta_{n})...S^{b"_{1},a"_{2}}
_{b"_{2},a"_{2}}(2i\pi +\beta_{1}-\beta_{2})\nonumber\\
& &\ \ \ \ F^{m_{1},...,m_{k}}_{b"_{1},
...,a"_{n}}(2i\pi +\alpha_{1},...,\alpha_{k}| \beta_{1},...,\beta_{n}) \ \ \ .
\end{eqnarray}

Like the bulk case,the system of functional equations (27),(28),(29) permit 
the reconstruction of the form-factors of boundary massive integrable model.
In other word,the equations (27),(28),(29) can be considered as the starting 
point to construct the form-factors of massive integrable model with 
integrable boundary.In this direction,some works have been given$^{[26]}$.
But the method of bosonization become more powerful and efficient.We will 
calculate the form-factors directly from the expressions of $|Lambda_{s}$ ,
$Z_{a}$ and $\langle B| $ ,$|B\rangle$ via q-oscillators.In this way, as 
we already prove,the main functional equations (27),(28),(29) are the 
direct results of 
the exchange relations (5) ,(25) , reflection properties (11) , (15) 
and the singular properties (8).

The procedure is as folows.We have obtain the bosonic representation 
of the operators $Z_{a}(\beta )$ , $\Lambda_{m}(\alpha )$ and $\Psi_{\pm }$ ,
which make (5) ,(8) ,(11) ,(15) ,(25) satisfied.After substituting the 
bosonic representation of $\Lambda _{s} (\alpha) $ ,$Z_{a}(\beta)$ in (25a) 
(7) into (26),we find that the form-factors (26) (or solutions to equations 
(27) (28) (29) ) can be given by the combination of multiply contour integrals 
as the following
\begin{eqnarray}
& &F^{m_{1},...,m_{k}}_{a_{1},...,a_{n}}
(\alpha_{1},...,\alpha_{k}| \beta_{1},...,\beta_{n})=\{ \prod_{a\in A}
\oint_{C_{a}} \frac{d\gamma_{a}}{2\pi}(q^{\frac{1}{2}}\Delta (\beta_{a}-
\gamma_{a})-q^{-\frac{1}{2}})\prod^{n}_{a>j} \Delta (\beta_{j}-\gamma_{a}) 
\nonumber\\
& & \ \ \times \prod_{d\in I_{-1}} \oint_{C_{d}} \frac{d\delta^{1}_{d} 
\delta^{2}_{d}}{4\pi ^{2}}[q'\Delta'(\alpha_{d}-\delta^{1}_{d})
\Delta'(\alpha_{d}-\delta^{2}_{d})-(q'+q'^{-1})\Delta '(\alpha_{d}-
\delta^{1}_{d})-q'^{-1}]\prod^{k}_{d>j} \Delta'(\alpha_{j}-
\delta^{1}_{d}) \Delta'(\alpha_{j}-
\delta^{2}_{d}) \nonumber\\
& & \ \ \times \prod_{b\in I_{0}} \oint_{C_{b}}\frac{d\delta_{b}}{2\pi}
(q'\Delta '(\alpha_{b}-
\delta_{b})-q'^{-1})\prod^{k}_{b>j} \Delta '(\alpha_{j}-\delta_{b}) \}
\nonumber\\
& & \ \ \times  R( \alpha_{1} .... \alpha_{k} | \delta^{1}_{1},\delta^{2}_{1},....
\delta^{1}_{l},\delta^{2}_{l},\delta_{1},....\delta_{t} | \beta_{1}
 .... \beta_{n}|\gamma_{1}....\gamma_{r})
\end{eqnarray}
where integral contour $C_{a}$,$C_{b}$,$C_{d}$ are taken by the rule in the 
sect.2,and 
\begin{eqnarray*}
& &\Delta(\beta)=\frac{\Gamma(-\frac{1}{2\xi}+\frac{i\beta}{\pi \xi})\Gamma(1+
\frac{1}{2\xi}-\frac{i\beta}{\pi \xi})}{\Gamma(-\frac{1}{2\xi}-
\frac{i\beta}{\pi \xi})\Gamma(1+\frac{1}{2\xi}+\frac{i\beta}{\pi \xi})}\\
& &\Delta'(\beta)=\frac{\Gamma(\frac{1}{2(1+\xi)}+\frac{i\beta}{\pi(1+\xi)})
\Gamma(1-\frac{1}{2(1+\xi)}-\frac{i\beta}{\pi(1+\xi)})}{\Gamma(\frac{1}
{2(1+\xi)}-\frac{i\beta}{\pi(1+\xi)})\Gamma(1-\frac{1}{2(1+\xi)}-
\frac{i\beta}{\pi(1+\xi)})}\\
& &A=\{ \ \ a_{j}|a_{j}=- \ \ \} \ \ \ , \ \ \ I_{0}=\{ \ \ m_{j}|m_{j}=0 \ \ \} \ \ \ , 
\ \ \ I_{-1}=\{ \ \ m_{j}|m_{j}=-1 \ \ \} 
\end{eqnarray*}
and the function $R( \alpha_{1} .... \alpha_{k} | \delta_{1} ....
\delta_{p} | \beta_{1} .... \beta_{n}|\gamma_{1}....\gamma_{r})$ are 
function like
\begin{eqnarray}
& &R( \alpha_{1} .... \alpha_{k} | \delta_{1} ....\delta_{p} | \beta_{1}
 .... \beta_{n}|\gamma_{1}....\gamma_{r}) \nonumber \\
& & \ \ \ =\frac{\langle0|e^{\Psi_{+}}\stackrel{\sim}{V}(\alpha_{k})....
\stackrel{\sim}{V}(\alpha_{1})\overline{V}^{'}(\delta_{p})
....\overline{V}^{'}(\delta_{1})V(\beta_{n})....V
(\beta_{1})\overline{V}(\gamma_{r})....\overline{V}
(\gamma_{1})e^{\Psi_{-}}|0\rangle}{\langle0|e^{\Psi_{+}}e^{\Psi_{-}}|0\rangle} \ \ .
\end{eqnarray}

Since the boundary operators $\Psi_{\pm}$ can not be 
explicitely expressed in the term of bosonic field $\phi (\beta)$ and 
$\phi'(\beta)$,it is difficult to get the function 
$R( \alpha_{1} .... \alpha_{k} | \delta_{1} ....\delta_{p} | \beta_{1}
 .... \beta_{n}|\gamma_{1}....\gamma_{r})$.
Fortunately,it can be viewed 
as the limits of their regularized counterparts 
$R_{\epsilon}( \alpha_{1} .... \alpha_{k} | \delta_{1} ....\delta_{p} | \beta_{1}
 .... \beta_{n}|\gamma_{1}....\gamma_{r})$ which can be obtained by changing  
 $\stackrel{\sim}{V}(\alpha)$ , $\overline{V}'(\delta)$ , $V(\beta)$ ,
$\overline{V}(\gamma)$ , $\Psi_{\pm}$ to  
 $\stackrel{\sim}{V}_{\epsilon}(\alpha)$ , $\overline{V}'_{\epsilon}(\delta)$
 , $V_{\epsilon}(\beta)$ , $\overline{V}_{\epsilon}(\gamma)$ , $\Psi_{\epsilon\pm}$
in (26). Because $\stackrel{\sim}{V}_{\epsilon}(\alpha)$ , $\overline{V}'_{\epsilon}
(\delta)$ , $V_{\epsilon}(\beta)$ , $\overline{V}_{\epsilon}(\gamma)$ , 
$\Psi_{\epsilon\pm}$ can be expanded into the q-oscillator (in sect.4),
we first move $e^{\Psi_{\epsilon -}}$ to the left of all vertex operators 
and reach the right side of $e^{\Psi_{\epsilon +}}$ .Then we insert a 
complete set of coheret states (38) between  $e^{\Psi_{\epsilon +}}$ and 
$e^{\Psi_{\epsilon -}}$ and we can obtain the exact form  
of function $R_{\epsilon}( \alpha_{1} .... \alpha_{k} | \delta_{1} ....
\delta_{p} | \beta_{1} .... \beta_{n}|\gamma_{1}....\gamma_{r})$ using the 
Gauss type integration formula (40)$^{[25]}$.
After taking 
the limit of $\epsilon \longrightarrow 0$ , we finally obtain the exact form of function 
$R( \alpha_{1} .... \alpha_{k} | \delta_{1} ....\delta_{p} | \beta_{1}
 .... \beta_{n}|\gamma_{1}....\gamma_{r})$ (see Appendix for details):
\begin{eqnarray*}
& &R(\alpha_{1}.... \alpha_{k}|\delta_{1}....\delta_{p}|\beta_{1}
....\beta_{n}|\gamma_{1}....\gamma_{r})=\delta_{k,p}\delta_{n,2r}
C_{1}^{r^{2}}C_{2}^{r}\\
& &\times \left\{\prod_{j=1}^{r}\overline{F}(\gamma_{j})\prod_{j<l}\overline{
G}(-\gamma_{j}-\gamma_{l})\overline{G}(\gamma_{j}-\gamma_{l})\right\}
\left\{\prod_{j=1}^{r}\prod_{l=1}^{n}W(\gamma_{j}-\beta_{l})W(-\gamma_{j}
-\beta_{l})\right\}\\
& &\times \left\{\prod_{j=1}^{r}\prod_{m=1}^{p}\overline{H}(\gamma_{j}-\delta_{m})
\overline{H}(-\gamma_{j}-\delta_{m})\right\}
\left\{\prod_{j=1}^{r}\prod_{n=1}^{k}\overline{H}^{-1}(\gamma_{j}-\alpha_{n})
\overline{H}^{-1}(-\gamma_{j}-\alpha_{n})\right\}\\
& &
\times \left\{\prod_{j=1}^{n}F(\beta_{j})\prod_{j<l}G(-\beta_{j}-\beta_{l})
G(\beta_{j}-\beta_{l})\right\}
\left\{\prod_{j=1}^{n}\prod_{l=1}^{p}U(\beta_{j}-\delta_{l})U(-\beta_{j}
-\delta_{l})\right\}\\
& &
\times  \left\{\prod_{j=1}^{n}\prod_{l=1}^{k}U^{-1}(\beta_{j}-\alpha_{l})
U^{-1}(-\beta_{j}-\alpha_{l})\right\} \left\{\prod_{j=1}^{p}\overline{F}^{'}
(\delta_{j})\prod_{j<l}\overline{G}^{'}(-\delta_{j}-\delta_{l})
\overline{G}^{'}(\delta_{j}-\delta_{l})\right\}\\
& &
\times \left\{\prod_{j=1}^{p}\prod_{l=1}^{k}\overline{G}^{'-1}(-\delta_{j}-
\alpha_{l})\overline{G}^{'-1}(\delta_{j}-\alpha_{l}) \right\}
\left\{\prod_{j=1}^{k}\stackrel{\sim }{F}(\alpha_{j})\prod_{j<l}\overline{G}^{'}
(-\alpha_{j}-\alpha_{l})\overline{G}^{'}(\alpha_{j}-\alpha_{l})\right\}\\
& &
\times \left\{\prod_{j=1}^{r}\overline{F}^{(0)}(\gamma_{j})\prod_{j=1}^{n}
F^{(0)}(\beta_{j})\prod_{j=1}^{p}\overline{F}^{'(0)}(\delta_{j})
\prod_{j=1}^{k}\stackrel{\sim }{F}^{(0)}(\alpha_{j}) \right\}
\end{eqnarray*}

where 
\begin{eqnarray*}
& & T (a_{1},a_{2},a_{3},a_{4},a_{5}|c_{1},c_{2},c_{3},c_{4}|b_{1},b_{2},b_{3})=
exp\{ \int_{0}^{\infty}\frac{sha_{1}xsha_{2}xsha_{3}xsha_{4}xsha_{5}xchb_{1}x}
{shc_{1}xshc_{2}xshc_{3}xshc_{4}x}\frac{e^{\frac{ix}{\pi}b_{2}}e^{-b_{3}x}}{x}dx\}\\ 
& &\overline{F}(\beta)=\overline{\rho}\overline{g}^{\frac{1}{2}}(-2\beta)
T ^{-1}(2,\frac{1}{2},\frac{1+\xi}{2},\frac{i\beta +\pi }{\pi },1|
1,\frac{\xi}{2},1,1|0,\beta ,1) \\
& &\ \ \ \ \ T ^{-1}(\frac{1}{2},\frac{1+\xi}{2},\frac{\beta}
{i\pi },\frac{\beta}{i\pi},1|1,\frac{\xi}{2},1,1|2,0,1)
T ^{-1}(\frac{1+\xi}{2},\frac{\beta}
{i\pi },\frac{\beta}{i\pi},1,1|\frac{\xi}{2},\frac{1}{2},1,1|0,0,1)\\
& &F(\beta)=\rho g^{\frac{1}{2}}(-2\beta)
T (\frac{1}{2},\frac{1+\xi}{2},\frac{i\beta +\pi }{\pi },1,1|
1,\frac{\xi}{2},1,1|0,\beta ,1) \\
& & \ \ \ \ \ T (\frac{1}{2},\frac{1+\xi}{2},\frac{\beta}
{i\pi },\frac{\beta}{i\pi},1|1,\frac{\xi}{2},1,1|0,0,1)
T (\frac{1}{2},\frac{1+\xi}{2},\frac{\beta}
{i\pi },\frac{\beta}{i\pi},1|1,\frac{\xi}{2},2,1|2,0,1)\\
& &\overline{F}^{'}(\beta)=\overline{\rho}^{'}\overline{g}^{'\frac{1}{2}}
(-2\beta)
T (\frac{1}{2},2,\frac{1+\xi}{2},\frac{i\beta +\pi }{\pi },\xi |
1,\frac{\xi}{2},1,1+\xi |0,\beta ,1)\\
& & \ \ \ \ \  T ^{-1}(\frac{\xi}{2},\frac{\beta}
{i\pi },\frac{\beta}{i\pi},1,1|\frac{1+\xi}{2},\frac{1}{2},1,1|0,0,1)
T (\frac{1}{2},\frac{1+\xi}{2},\frac{\beta}
{i\pi },\frac{\beta}{i\pi},\xi |1,\frac{\xi}{2},1,1+\xi |2,0,1)\\
& &\stackrel{\sim}{F}(\beta)=\overline{\rho}^{'} 
g^{'\frac{1}{2}}(-2\beta)
T ^{-1}(\frac{1}{2},2,\frac{1+\xi}{2},\frac{i\beta +\pi }{\pi },\xi |
1,\frac{\xi}{2},1,1+\xi |0,\beta ,1)\\
& & \ \ \ \ \  T ^{-1}(\frac{\xi}{2},\frac{\beta}
{i\pi },\frac{\beta}{i\pi},1,1|\frac{1+\xi}{2},\frac{1}{2},1,1|0,0,1)
T (\frac{1}{2},\frac{1+\xi}{2},\frac{\beta}
{i\pi },\frac{\beta}{i\pi},\xi |1,\frac{\xi}{2},1,1+\xi |2,0,1)\\
& &\overline{G}(\beta)=\overline{g}(\beta)
T ^{-2}(\frac{1+\xi}{2},\frac{\beta}
{2i\pi },\frac{\beta}{2i\pi},1,1|\frac{\xi}{2},\frac{1}{2},1,1|0,0,1)\\
& &G(\beta)=g(\beta)
T ^{-2}(\frac{1}{2},\frac{1+\xi}{2},\frac{\beta}
{2i\pi },\frac{\beta}{2i\pi},1|\frac{\xi}{2},1,1,1|0,0,1)\\
& &\overline{G}^{'}(\beta)=\overline{g}^{'}(\beta)
T ^{-2}(\frac{\xi}{2},\frac{\beta}
{2i\pi },\frac{\beta}{2i\pi},1,1|\frac{1+\xi}{2},\frac{1}{2},1,1|0,0,1)\\
& &W(\beta)=w(\beta)
T (\frac{1+\xi}{2},\frac{\beta}
{2i\pi },\frac{\beta}{2i\pi},1,1|\frac{\xi}{2},1,1,1|0,0,1)\\
& &\overline{H}(\beta)=\overline{h}(\beta)
T ^{-2}(\frac{\beta}
{2i\pi },\frac{\beta}{2i\pi},1,1,1|\frac{1}{2},1,1,1|0,0,1)
\ \ ,\ \ U(\beta)=u(\beta)
T ^{-2}(\frac{\beta}
{2i\pi },\frac{\beta}{2i\pi},1,1,1|1,1,1,1|0,0,1)\\
& &w(\beta)=\frac{\Gamma(-\frac{1}{2\xi}+\frac{i\beta}{\pi\xi})}
{\Gamma(1+\frac{1}{2\xi}+\frac{i\beta}{\pi\xi})} 
\ \ , \ \ \mu(\beta)=\frac{i\beta}{2\pi}\ \ ,\ \ 
\overline{h}(\beta)=\frac{4\pi^{2}}{\beta^{2}+\frac{\pi^{2}}{4}}\\
& &g(\beta)=\left\{\frac{1}{\Gamma(\frac{1}{\xi})}\right\}^{\frac{1}{2}}
\frac{\Gamma(\frac{1}{\xi}+\frac{i\beta}{\pi\xi})}{\Gamma(\frac{i\beta}
{\pi\xi})}\prod_{p=1}^{\infty}\frac{[R_{p}(i\pi)R_{p}(0)]^{\frac{1}{2}}}{R_{p}(
\beta)} \\
& &\overline{g}(\beta)=\frac{i\beta\Gamma(1+\frac{1}{\xi}+\frac{i\beta}{\pi\xi})}
{\pi\xi \Gamma(-\frac{1}{\xi}+\frac{i\beta}{\pi\xi})}\ \ ,\ \ \ 
 \overline{g}^{'}(\beta)=
\frac{i\beta\Gamma(1-\frac{1}{1+\xi}+\frac{i\beta}{\pi(1+\xi)})}
{\pi(1+\xi) \Gamma(\frac{1}{1+\xi}+\frac{i\beta}{\pi(1+\xi)})}\\
& &\rho ^{2}=\frac{i}{\pi\xi}[\Gamma(\frac{1}{\xi})]^{\frac{1}{2}}\prod_{p=1}^{
\infty}[\frac{R_{p}(i\pi)}{R_{p}(0)}]^{\frac{1}{2}}\ \ ,\ \ 
\overline{\rho}^{2}=\frac{i\Gamma(1+\frac{1}{\xi})}{\pi\xi \Gamma(-
\frac{1}{\xi})} \ \ , \ \ \overline{\rho}^{'2}=\frac{i\Gamma(1-\frac{1}{
1+\xi})}{\pi(1+\xi)\Gamma(\frac{1}{1+\xi})} \\
& &\overline{F}^{(0)}(\beta)=T ^{2}(\frac{2i\mu}{\pi}-1-\xi ,\frac{i\beta 
+\pi}{\pi},1,1,1|\xi ,1,1,1|0,\beta ,1)
T ^{2}(\frac{3}{2},\frac{1-\xi }{2},\frac{i\beta 
+\pi}{\pi},1,1|\frac{\xi }{2},1,1,1|0,\beta ,1)\\
& &\ \ \ \ T ^{2}(\frac{2i\mu}{\pi}-1-\xi ,\frac{\beta }{i\pi},
\frac{i\beta }{\pi},1,1|\xi ,2,1,1|1,0,1)
T ^{2}(\frac{3}{2},\frac{1-\xi }{2},\frac{i\beta 
}{\pi},\frac{i\beta }{\pi},1|\frac{\xi }{2},2,1,1|1,0,1)\\
& &\ \ \ \ T ^{-1}(\xi +\frac{2i\mu}{\pi},\frac{\beta}{i\pi},\frac{\beta}{i\pi},
1,1|\xi ,1,1,1|0,0,0)\\
& &F^{(0)}(\beta)=T ^{-2}(\frac{2i\mu}{\pi}-1-\xi ,\frac{i\beta 
+\pi}{\pi},1,1,1|\xi ,2,1,1|0,\beta ,1)
T ^{-2}(\frac{3}{2},\frac{1-\xi }{2},\frac{i\beta 
+\pi}{\pi},1,1|\frac{\xi }{2},2,1,1|0,\beta ,1)\\
& &\ \ \ \ T ^{-2}(\frac{2i\mu}{\pi}-1-\xi ,\frac{\beta }{i\pi},
\frac{i\beta }{\pi},1,1|\xi ,2,2,1|1,0,1)
T ^{-2}(\frac{3}{2},\frac{1-\xi }{2},\frac{i\beta 
}{\pi},\frac{i\beta }{\pi},1|\frac{\xi }{2},2,2,1|1,0,1)\\
& &\ \ \ \ T (\xi +\frac{2i\mu}{\pi},\frac{\beta}{i\pi},\frac{\beta}{i\pi},
1,1|\xi ,2,1,1|0,0,0)\\
& &\overline{F}^{'(0)}(\beta)=T ^{2}(\frac{2i\mu}{\pi}-1-\xi ,\frac{i\beta 
+\pi}{\pi},1,1,1|1+\xi ,1,1,1|0,\beta ,1)
T ^{-2}(\frac{3}{2},\frac{1-\xi }{2},\xi,\frac{i\beta 
+\pi}{\pi},1|\frac{\xi }{2},1+\xi ,1,1|0,\beta ,1)\\
& &\ \ \ \ T ^{-2}(\frac{2i\mu}{\pi}-1-\xi ,\frac{\beta }{i\pi},
\frac{i\beta }{\pi},1,1|1+\xi ,2,1,1|1,0,1)
T ^{-2}(\frac{3}{2},\frac{1-\xi }{2},\xi ,\frac{i\beta 
}{\pi},\frac{i\beta }{\pi}|\frac{\xi }{2},2,1+\xi ,1|1,0,1)\\
& &\ \ \ \ T (\xi +\frac{2i\mu}{\pi},\frac{\beta}{i\pi},\frac{\beta}{i\pi},
1,1|1+\xi ,1,1,1|0,0,0)\\
& &\stackrel{\sim}{F}^{(0)}(\beta)=\{\overline{F}^{'(0)}(\beta)\}^{-1}\\
& &C_{1}=exp\{-\int_{0}^{\infty}\frac{sh\frac{x}{2}sh\frac{x(1+\xi )}{2}}{sh
\frac{x\xi }{2}}(\frac{1}{sh^{2}\frac{x}{2}}+\frac{4}{sh^{2}x}-\frac{4}{sh
\frac{x}{2}shx})\frac{e^{-x}}{x}dx \}\\
\end{eqnarray*}
\begin{eqnarray}
& &C_{2}=exp\{\int_{0}^{\infty}\frac{sh\frac{x}{2}sh\frac{x(1+\xi )}{2}(1-
\frac{sh2x}{2shx})}{shxsh\frac{x\xi }{2}}\frac{e^{-x}}{x}dx-\frac{1}{2}
\int_{0}^{\infty}\frac{sh\frac{x}{2}sh\frac{x(1+\xi )}{2}ch2x(\frac{1}{shx}-
\frac{2}{sh2x})}{shxsh\frac{x\xi }{2}}\frac{e^{-x}}{x}dx\nonumber\\
& &\ \ \ \ \ \ \ \ -\int_{0}^{\infty}\frac{shx(\frac{2i\mu}{\pi}-1-\xi )
(\frac{2shx}{sh2x}-1)}{
shx\xi }\frac{e^{-x}}{x}dx-\int_{0}^{\infty}\frac{sh\frac{3x}{2}sh\frac{x(1-\xi )
}{2}(\frac{2shx}{sh2x}-1)}{sh\frac{x\xi }{2}shx}\frac{e^{-x}}{x}dx\nonumber\\
& &\ \ \ \ \ \  +\int_{0}^{\infty}\frac{shx(\frac{2i\mu}{\pi}-1-\xi )chx(-\frac{2shx}{sh2x}+1)}{
shx\xi sh2x}\frac{e^{-x}}{x}dx+\int_{0}^{\infty}\frac{sh\frac{3x}{2}
sh\frac{x(1-\xi )}{2}chx(\frac{2shx}{sh2x}-1)}{sh\frac{x\xi }{2}shxsh2x}
\frac{e^{-x}}{x}dx\nonumber\\
& &\ \ \ \ \ \ \ \  -\frac{1}{2}\int_{0}^{\infty}\frac{shx(\xi +\frac
{2i\mu}{\pi})(1-\frac{2shx}
{sh2x})}{shx\xi shx}\frac{dx}{x}\} \ \ .
\end{eqnarray}

Therefore,we obtained the form-factors of SG model with a fixed boundary 
condition,which is expressed in terms of integrals ((31)).This is quite 
similar to the correlation functions in CFT$^{[19,20]}$ and the 
form-factors in 
massive integrable model without any boundary$^{[10,35]}$,they are also 
expressed in terms of integrals.
\section{Discussion}
Let us summarize our resluts.We have shown that the obtained 
form factors of semi-infinite line SG theory with a fixed boundary obey 
the quantum boundary KZ equation( (27),(28),(29) ).The quantum KZ equation is a simple 
consequence of the exchange relation of ZF algebra and the reflection 
properties (11),(15) of the boundary ground state and its dual.It is believed 
that these quantum difference relations should govern the form-factors
 in the integrable field theories with integrable boundary as 
the classical KZ equation does in CFT.

In this paper,we only consider SG theory with fixed boundary condition which 
corresponds to the diagonal solution to the boundary Yang-Baxter equation.To 
solve SG theories with other generic integrable boundary condition is a 
challenging work;When the coupling constant $\xi $ is equal to some special 
value ($\xi =\frac{1}{N}$) ,which may be related to th Restricted SG model,
it is worthy of further studying .We hope to present the full results 
eslewhere. The 
connection between the symmetries of SG model and deformed Virasoro algebra 
is also worth of investigating.

\section*{Acknowledgement} 

 We wish to thank Prof. M. Jimbo ,V.E. Korepin ,P.B. Wiegmann and 
Dr. L. Chao for  fruitful 
discussions.This work is supported 
in part by the National Natural Science Fund of China. 

\section*{Appendix}
\subsection*{A.1}
In this appendix, we first calculate the regularized function 
$R_{\epsilon}( \alpha_{1} .... \alpha_{k} | \delta_{1} ....\delta_{p} |
 \beta_{1} .... \beta_{n}|\gamma_{1}....\gamma_{r})$ using 
the q-oscillator expansion of vertex operators and the boundary operators 
in sect.4 .Then taking the limit 
of $\epsilon \longrightarrow 0$ ,we obtain the function 
$R( \alpha_{1} .... \alpha_{k} | \delta_{1} ....\delta_{p} |
 \beta_{1} .... \beta_{n}|\gamma_{1}....\gamma_{r})$ .

Using the exact form of $\Psi _{\epsilon -}$ in (23) ,one can find that 
$e^{\Psi _{\epsilon -}}$ has the effect of a Bogoliubov transformation 
\begin{eqnarray}
& &e^{-\Psi_{\epsilon -}}a_{-n}e^{\Psi_{\epsilon -}}=a_{-n} \ \ \ \ ,\ \ \ n>0 
\nonumber\\
& &e^{-\Psi_{\epsilon -}}a_{n}e^{\Psi_{\epsilon -}}=a_{n}-a_{-n} +\lambda _{m} \ \ \ ,
\ \ \ \ n>0\ \ \ .
\end{eqnarray}
From (21) and (34),one can get the reflecting properties
\begin{eqnarray}
& &e^{i\phi_{\epsilon}(\beta)}|B\rangle =\rho _{\epsilon}exp\{-\frac{1}{2}
\sum^{\infty}_{m=1}\frac{sh\frac{\pi m\epsilon }{2}
sh\frac{\pi m\epsilon (1+\xi )}{2}}{msh\pi m\epsilon
sh\frac{\pi m\epsilon \xi }{2}}e^{2im\epsilon \beta}
+\sum ^{\infty}_{m=1}\frac{\lambda _{m}}{sh\pi m\epsilon}
e^{im\epsilon \beta}\}e^{i\phi_{\epsilon -}(-\beta)+
i\phi_{\epsilon -}(\beta)}|B\rangle \nonumber\\
& &e^{-i\overline{\phi}_{\epsilon}(\beta)}|B\rangle =\overline{\rho }
_{\epsilon}exp\{-\frac{1}{2}
\sum^{\infty}_{m=1}\frac{sh\pi m\epsilon 
sh\frac{\pi m\epsilon (1+\xi )}{2}}{msh\frac{\pi m\epsilon }{2}
sh\frac{\pi m\epsilon \xi }{2}}e^{2im\epsilon \beta}
-\sum ^{\infty}_{m=1}\frac{\lambda _{m}}{sh\frac{\pi m\epsilon}{2}}
e^{im\epsilon \beta}\}e^{-i\overline{\phi}_{\epsilon -}(-\beta)-
i\overline{\phi}_{\epsilon -}(\beta)}|B\rangle \nonumber\\
& &e^{-i\overline{\phi}^{'}_{\epsilon}(\beta)}|B\rangle =\overline{\rho }^{'}
_{\epsilon}exp\{-\frac{1}{2}
\sum^{\infty}_{m=1}\frac{sh\pi m\epsilon 
sh\frac{\pi m\epsilon \xi }{2}}{msh\frac{\pi m\epsilon }{2}
sh\frac{\pi m\epsilon (1+\xi )}{2}}e^{2im\epsilon \beta}
+\sum ^{\infty}_{m=1}\frac{sh\frac{\pi m\epsilon \xi }{2} \ \ \lambda _{m}}
{sh\frac{\pi m\epsilon }{2}sh\frac{\pi m\epsilon (1+\xi )}{2}}
e^{im\epsilon \beta}\}e^{-i\overline{\phi}^{'}_{\epsilon -}(-\beta)-
i\overline{\phi}^{'}_{\epsilon -}(\beta)}|B\rangle \nonumber\\
& &e^{i\overline{\phi}^{'}_{\epsilon}(\beta)}|B\rangle =\overline{\rho }^{'}
_{\epsilon}exp\{-\frac{1}{2}
\sum^{\infty}_{m=1}\frac{sh\pi m\epsilon 
sh\frac{\pi m\epsilon \xi }{2}}{msh\frac{\pi m\epsilon }{2}
sh\frac{\pi m\epsilon (1+\xi )}{2}}e^{2im\epsilon \beta}
-\sum ^{\infty}_{m=1}\frac{sh\frac{\pi m\epsilon \xi }{2} \ \ \lambda _{m}}
{sh\frac{\pi m\epsilon }{2}sh\frac{\pi m\epsilon (1+\xi )}{2}}
e^{im\epsilon \beta}\}e^{i\overline{\phi}^{'}_{\epsilon -}(-\beta)+
i\overline{\phi}^{'}_{\epsilon -}(\beta)}|B\rangle . \nonumber\\
& & \ \ 
\end{eqnarray}
Noting that the operators 
$\stackrel{\sim}{V}_{\epsilon}(\alpha)$ , $\overline{V}'_{\epsilon}
(\delta)$ , $V_{\epsilon}(\beta)$ , $\overline{V}_{\epsilon}(\gamma)$ 
have the following normal-ordering$^{[12]}$
\begin{eqnarray*}
& &V_{\epsilon}(\beta _{2})V_{\epsilon}(\beta _{1})=\rho_{\epsilon}^{2}
g_{\epsilon}(\beta_{1}-\beta_{2}):
V_{\epsilon}(\beta _{2})V_{\epsilon}(\beta _{1}):\\
& &V_{\epsilon}(\beta _{2})\overline{V}_{\epsilon}(\gamma _{1})=
\rho_{\epsilon}\overline{\rho}_{\epsilon}
w_{\epsilon}(\gamma_{1}-\beta_{2}):
V_{\epsilon}(\beta _{2})\overline{V}_{\epsilon}(\gamma _{1}):\\
& &\overline{V}_{\epsilon}(\gamma _{2})\overline{V}_{\epsilon}(\gamma _{1})=
\overline{\rho}_{\epsilon}^{2}
\overline{g}_{\epsilon}(\gamma_{1}-\gamma_{2}):
\overline{V}_{\epsilon}(\gamma _{2})\overline{V}_{\epsilon}(\gamma _{1}):\\
& &\overline{V}'_{\epsilon}(\delta _{2})\overline{V}'_{\epsilon}(\delta _{1})=
\overline{\rho}_{\epsilon}^{'2}
\overline{g}^{'}_{\epsilon}(\delta_{1}-\delta_{2}):
\overline{V}'_{\epsilon}(\delta _{2})\overline{V}_{\epsilon}(\delta _{1}):\\
& &\overline{V}'_{\epsilon}(\delta _{2})\stackrel{\sim}{V}'_{\epsilon}
(\alpha _{1})=\overline{\rho}_{\epsilon}^{'2}
\overline{g}^{'-1}_{\epsilon}(\alpha_{1}-\delta_{2}):
\overline{V}'_{\epsilon}(\delta _{2})\stackrel{\sim}{V}_{\epsilon}
(\alpha_{1}):\\
& &\stackrel{\sim}{V}_{\epsilon}(\alpha _{2})\stackrel{\sim}{V}_{\epsilon}
(\alpha _{1})=\overline{\rho}_{\epsilon}^{'2}
\overline{g}^{'}_{\epsilon}(\alpha_{1}-\alpha_{2}):
\stackrel{\sim}{V}_{\epsilon}(\alpha_{2})\stackrel{\sim}{V}_{\epsilon}
(\alpha_{1}):\\
& &V_{\epsilon }(\beta _{2})\overline{V}'_{\epsilon }
(\delta _{1})=\overline{\rho}^{'}_{\epsilon }\rho _{\epsilon}
\mu _{\epsilon}(\delta_{1}-\beta_{2}):
V_{\epsilon}(\beta_{2})\overline{V}'_{\epsilon}
(\delta _{1}):\\
& &V_{\epsilon}(\beta _{2})\stackrel{\sim}{V}_{\epsilon}
(\alpha _{1})=\overline{\rho}^{'}_{\epsilon}\rho_{\epsilon}
\mu^{-1}_{\epsilon}(\alpha_{1}-\beta_{2}):
V_{\epsilon}(\beta _{2})\stackrel{\sim}{V}_{\epsilon}
(\alpha_{1}):\\
& &\overline{V}_{\epsilon}(\gamma _{2})\overline{V}'_{\epsilon}
(\delta _{1})=\overline{\rho}^{'}_{\epsilon}\overline{\rho}_{\epsilon}
\overline{h}_{\epsilon}(\delta_{1}-\gamma_{2}):
\overline{V}_{\epsilon}(\gamma_{2})\overline{V}'_{\epsilon}
(\delta_{1}):\\
& &\overline{V}_{\epsilon}(\gamma _{2})\stackrel{\sim}{V}_{\epsilon}
(\alpha_{1})=\overline{\rho}^{'}_{\epsilon}\overline{\rho}_{\epsilon}
\overline{h}^{-1}_{\epsilon}(\alpha_{1}-\gamma_{2}):
\overline{V}_{\epsilon}(\gamma_{2})\stackrel{\sim}{V}_{\epsilon}
(\alpha_{1}):\\
\end{eqnarray*}
where functions $g_{\epsilon}(\beta)$ ,$\overline{g}_{\epsilon}(\beta)$ ,
$w_{\epsilon}(\beta)$ ,$\overline{g}^{'}_{\epsilon}(\beta)$ ,
$\overline{h}_{\epsilon}(\beta)$ ,$\mu_{\epsilon}(\beta)$ are the regularized version 
of $g(\beta)$ ,$\overline{g}(\beta)$ ,$w(\beta)$ ,$\overline{g}^{'}(\beta)$ ,
$\overline{h}(\beta)$ ,$\mu(\beta)$ ,and using  (35),
one can move $e^{\Psi_{-}}$ to the left of all vertex operators and obtain 
\begin{eqnarray}
& &R_{\epsilon}( \alpha_{1} .... \alpha_{k} | \delta_{1} ....\delta_{p} | \beta_{1}
 .... \beta_{n}|\gamma_{1}....\gamma_{r}) \nonumber \\
& & \ \ =\frac{\langle 0|e^{\Psi_{\epsilon +}}\stackrel{\sim}{V}_{\epsilon}
(\alpha_{k})....
\stackrel{\sim}{V}_{\epsilon}(\alpha_{1})\overline{V}^{'}_{\epsilon}
(\delta_{p})
....\overline{V}^{'}_{\epsilon}(\delta_{1})V_{\epsilon}(\beta_{n})....
V_{\epsilon}(\beta_{1})\overline{V}_{\epsilon}(\gamma_{r})....
\overline{V}_{\epsilon}
(\gamma_{1})e^{\Psi_{\epsilon -}}|0\rangle }{\langle 0|e^{\Psi_{\epsilon +}}
e^{\Psi_{\epsilon -}}|0\rangle } \nonumber\\
& &\ \ \ =\delta_{k,p}\delta_{n,2r}
\left\{\prod_{j=1}^{r}\overline{f}_{\epsilon}(\gamma_{j})\prod_{j<l}
\overline{g}_{\epsilon}(-\gamma_{j}-\gamma_{l})\overline{g}_{\epsilon}
(\gamma_{j}-\gamma_{l})\right\}
\left\{\prod_{j=1}^{r}\prod_{l=1}^{n}w_{\epsilon}(\gamma_{j}-\beta_{l})
w_{\epsilon}(-\gamma_{j}
-\beta_{l})\right\}\nonumber\\
& &\ \ \ \times \left\{\prod_{j=1}^{r}\prod_{m=1}^{p}\overline{h}_{\epsilon}
(\gamma_{j}-\delta_{m})
\overline{h}_{\epsilon}(-\gamma_{j}-\delta_{m})\right\}
\left\{\prod_{j=1}^{r}\prod_{n=1}^{k}\overline{h}_{\epsilon}^{-1}
(\gamma_{j}-\alpha_{n})
\overline{h}_{\epsilon}^{-1}(-\gamma_{j}-\alpha_{n})\right\}\nonumber\\
& &\ \ \ 
\times \left\{\prod_{j=1}^{n}f_{\epsilon}(\beta_{j})\prod_{j<l}
g_{\epsilon}(-\beta_{j}-\beta_{l})
g_{\epsilon}(\beta_{j}-\beta_{l})\right\}
\left\{\prod_{j=1}^{n}\prod_{l=1}^{p}\mu_{\epsilon}(\beta_{j}-\delta_{l})
\mu_{\epsilon}(-\beta_{j}
-\delta_{l})\right\}\nonumber\\
& &\ \ \ 
\times  \left\{\prod_{j=1}^{n}\prod_{l=1}^{k}\mu_{\epsilon}^{-1}
(\beta_{j}-\alpha_{l})
\mu_{\epsilon}^{-1}(-\beta_{j}-\alpha_{l})\right\} 
\left\{\prod_{j=1}^{p}\overline{f}^{'}_{\epsilon}
(\delta_{j})\prod_{j<l}\overline{g}^{'}_{\epsilon}(-\delta_{j}-\delta_{l})
\overline{g}^{'}_{\epsilon}(\delta_{j}-\delta_{l})\right\}\nonumber\\
& &\ \ \ 
\times \left\{\prod_{j=1}^{p}\prod_{l=1}^{k}\overline{g}^{'-1}_{\epsilon}
(-\delta_{j}-
\alpha_{l})\overline{g}^{'-1}_{\epsilon}(\delta_{j}-\alpha_{l}) \right\}
\left\{\prod_{j=1}^{k}\stackrel{\sim }{f}_{\epsilon}(\alpha_{j})
\prod_{j<l}\overline{g}^{'}_{\epsilon}
(-\alpha_{j}-\alpha_{l})\overline{g}^{'}_{\epsilon}
(\alpha_{j}-\alpha_{l})\right\}\nonumber\\
& &\ \ \ \times I( \alpha_{1} .... \alpha_{k} |
 \delta_{1} ....\delta_{p} |
 \beta_{1} .... \beta_{n}|\gamma_{1}....\gamma_{r})
\end{eqnarray}
where
\begin{eqnarray*}
& &I( \alpha_{1} .... \alpha_{k} |
 \delta_{1} ....\delta_{p} |
 \beta_{1} .... \beta_{n}|\gamma_{1}....\gamma_{r})\\
& & \ \ \ =\frac{\langle0|e^{\Psi_{\epsilon +}}e^{\Psi_{\epsilon -}}exp\{\sum 
^{\infty}_{m=1} a_{-m}(X_{m}-Y_{m})\}|0\rangle }
{\langle0|e^{\Psi_{\epsilon +}}e^{\Psi_{\epsilon -}}|0\rangle },\\
& & \ \ \ X_{m}=\frac{\sum ^{r}_{j=1}e^{-im\epsilon \gamma_{j}}}
{sh\frac{\pi m\epsilon}{2}}
-\frac{\sum ^{n}_{l=1}e^{-im\epsilon \beta_{l}}}
{sh\pi m\epsilon}
-\frac{sh\frac{\pi m\epsilon \xi}{2}\sum ^{p}_{l=1}e^{-im\epsilon \delta_{l}}}
{sh\frac{\pi m\epsilon}{2}sh\frac{\pi m\epsilon (1+\xi)}{2}}
+\frac{sh\frac{\pi m\epsilon \xi}{2}\sum ^{k}_{l=1}e^{-im\epsilon \alpha_{l}}}
{sh\frac{\pi m\epsilon}{2}sh\frac{\pi m\epsilon (1+\xi)}{2}},\\
& &\ \ \ \ Y_{m}=-\frac{\sum ^{r}_{j=1}e^{im\epsilon \gamma_{j}}}
{sh\frac{\pi m\epsilon}{2}}
+\frac{\sum ^{n}_{l=1}e^{im\epsilon \beta_{l}}}
{sh\pi m\epsilon}
+\frac{sh\frac{\pi m\epsilon \xi}{2}\sum ^{p}_{l=1}e^{im\epsilon \delta_{l}}}
{sh\frac{\pi m\epsilon}{2}sh\frac{\pi m\epsilon (1+\xi)}{2}}
+\frac{sh\frac{\pi m\epsilon \xi}{2}\sum ^{k}_{l=1}e^{im\epsilon \alpha_{l}}}
{sh\frac{\pi m\epsilon}{2}sh\frac{\pi m\epsilon (1+\xi)}{2}} \ \ .
\end{eqnarray*} 
We then insert between $e^{\Psi_{\epsilon +}}$ and $e^{\Psi_{\epsilon -}}$ the 
completeness relation of the coherent states (39) and use the 
the integration formula (40)$^{[25]}$. As a result, we have 
\begin{eqnarray}
& &I( \alpha_{1} .... \alpha_{k} |
 \delta_{1} ....\delta_{p} |
 \beta_{1} .... \beta_{n}|\gamma_{1}....\gamma_{r})\nonumber\\
& & \ \ \ \ =exp\{\frac{1}{2}\sum ^{\infty}_{m=1}[-
\frac{sh\frac{\pi m\epsilon}{2}sh\frac{\pi m\epsilon (1+\xi)}{2}
e^{-\pi m\epsilon}}{2msh\frac{\pi m\epsilon \xi}{2}}(X^{2}_{m}
+Y^{2}_{m}-2X_{m}Y_{m})\nonumber\\
& &\ \ \ \ +\frac{1}{1-e^{-2\pi m\epsilon}}(\rho_{m}
-e^{-2\pi m\epsilon}\lambda_{m})(X_{m}-Y_{m})]\}\ \ \ .
\end{eqnarray}
Substituting eq.(37) into (36) ,we find the form of regularized 
function $ R_{\epsilon}( \alpha_{1} .... \alpha_{k} |
 \delta_{1} ....\delta_{p} | \beta_{1}
 .... \beta_{n}|\gamma_{1}....\gamma_{r})$ is 
\begin{eqnarray}
& &R_{\epsilon}( \alpha_{1} .... \alpha_{k} | \delta_{1} ....\delta_{p} | \beta_{1}
 .... \beta_{n}|\gamma_{1}....\gamma_{r}) \nonumber \\
& &\ \ \ =C^{r^{2}}_{\epsilon 1}C^{r}_{\epsilon 2}
\delta_{k,p}\delta_{n,2r}
\left\{\prod_{j=1}^{r}\overline{F}_{\epsilon}(\gamma_{j})\prod_{j<l}
\overline{G}_{\epsilon}(-\gamma_{j}-\gamma_{l})\overline{G}_{\epsilon}
(\gamma_{j}-\gamma_{l})\right\}
\left\{\prod_{j=1}^{r}\prod_{l=1}^{n}W_{\epsilon}(\gamma_{j}-\beta_{l})
W_{\epsilon}(-\gamma_{j}
-\beta_{l})\right\}\nonumber\\
& &\ \ \ \times \left\{\prod_{j=1}^{r}\prod_{m=1}^{p}\overline{H}_{\epsilon}
(\gamma_{j}-\delta_{m})
\overline{H}_{\epsilon}(-\gamma_{j}-\delta_{m})\right\}
\left\{\prod_{j=1}^{r}\prod_{n=1}^{k}\overline{H}_{\epsilon}^{-1}
(\gamma_{j}-\alpha_{n})
\overline{H}_{\epsilon}^{-1}(-\gamma_{j}-\alpha_{n})\right\}\nonumber\\
& &\ \ \ 
\times \left\{\prod_{j=1}^{n}f_{\epsilon}(\beta_{j})\prod_{j<l}
G_{\epsilon}(-\beta_{j}-\beta_{l})
G_{\epsilon}(\beta_{j}-\beta_{l})\right\}
\left\{\prod_{j=1}^{n}\prod_{l=1}^{p}U_{\epsilon}(\beta_{j}-\delta_{l})
U_{\epsilon}(-\beta_{j}
-\delta_{l})\right\}\nonumber\\
& &\ \ \ 
\times  \left\{\prod_{j=1}^{n}\prod_{l=1}^{k}U_{\epsilon}^{-1}
(\beta_{j}-\alpha_{l})
U_{\epsilon}^{-1}(-\beta_{j}-\alpha_{l})\right\} 
\left\{\prod_{j=1}^{p}\overline{F}^{'}_{\epsilon}
(\delta_{j})\prod_{j<l}\overline{G}^{'}_{\epsilon}(-\delta_{j}-\delta_{l})
\overline{G}^{'}_{\epsilon}(\delta_{j}-\delta_{l})\right\}\nonumber\\
& &\ \ \ 
\times \left\{\prod_{j=1}^{p}\prod_{l=1}^{k}\overline{G}^{'-1}_{\epsilon}
(-\delta_{j}-
\alpha_{l})\overline{G}^{'-1}_{\epsilon}(\delta_{j}-\alpha_{l}) \right\}
\left\{\prod_{j=1}^{k}\stackrel{\sim }{F}_{\epsilon}(\alpha_{j})
\prod_{j<l}\overline{G}^{'}_{\epsilon}
(-\alpha_{j}-\alpha_{l})\overline{G}^{'}_{\epsilon}
(\alpha_{j}-\alpha_{l})\right\}\nonumber\\
& &\ \ \ 
\times \left\{\prod_{j=1}^{r}\overline{F}^{(0)}_{\epsilon}
(\gamma_{j})\prod_{j=1}^{n}
F^{(0)}_{\epsilon}(\beta_{j})\prod_{j=1}^{p}\overline{F}^{'(0)}_{\epsilon}
(\delta_{j})
\prod_{j=1}^{k}\stackrel{\sim }{F}^{(0)}_{\epsilon}(\alpha_{j}) \right\}\ \ \ ,\nonumber
\end{eqnarray}
where 
\begin{eqnarray*}
& & T_{\epsilon}(a_{1},a_{2},a_{3},a_{4},a_{5}|c_{1},c_{2},c_{3},c_{4}
|b_{1},b_{2},b_{3})\\
& &\ \ \ \ =exp\left\{ 
\sum^{\infty}_{m=1}\frac{sha_{1}m\epsilon sha_{2}m\epsilon sha_{3}m\epsilon 
sha_{4}m\epsilon sha_{5}m\epsilon chb_{1}m\epsilon 
e^{\frac{im\epsilon b_{2}}{\pi}}e^{-b_{3}m\epsilon}}{mshc_{1}m\epsilon 
shc_{2}m\epsilon shc_{3}m\epsilon shc_{4}m\epsilon }\right\}\\
& &\overline{F}_{\epsilon}(\beta)=\overline{\rho}_{\epsilon}
\overline{g}_{\epsilon}^{\frac{1}{2}}(-2\beta)
T_{\epsilon}^{-1}(2,\frac{1}{2},\frac{1+\xi}{2},\frac{i\beta +\pi }{\pi },1|
1,\frac{\xi}{2},1,1|0,\beta ,1) \\
& &\ \ \ \ \ T_{\epsilon}^{-1}(\frac{1}{2},\frac{1+\xi}{2},\frac{\beta}
{i\pi },\frac{\beta}{i\pi},1|1,\frac{\xi}{2},1,1|2,0,1)
T_{\epsilon}^{-1}(\frac{1+\xi}{2},\frac{\beta}
{i\pi },\frac{\beta}{i\pi},1,1|\frac{\xi}{2},\frac{1}{2},1,1|0,0,1)\\
& &F_{\epsilon}(\beta)=\rho_{\epsilon} g_{\epsilon}^{\frac{1}{2}}(-2\beta)
T_{\epsilon}(\frac{1}{2},\frac{1+\xi}{2},\frac{i\beta +\pi }{\pi },1,1|
1,\frac{\xi}{2},1,1|0,\beta ,1) \\
& & \ \ \ \ \ T_{\epsilon}(\frac{1}{2},\frac{1+\xi}{2},\frac{\beta}
{i\pi },\frac{\beta}{i\pi},1|1,\frac{\xi}{2},1,1|0,0,1)
T_{\epsilon}(\frac{1}{2},\frac{1+\xi}{2},\frac{\beta}
{i\pi },\frac{\beta}{i\pi},1|1,\frac{\xi}{2},2,1|2,0,1)\\
& &\overline{F}_{\epsilon}^{'}(\beta)=\overline{\rho}_{\epsilon}^{'}
\overline{g}_{\epsilon}^{'\frac{1}{2}}
(-2\beta)
T_{\epsilon}(\frac{1}{2},2,\frac{1+\xi}{2},\frac{i\beta +\pi }{\pi },\xi |
1,\frac{\xi}{2},1,1+\xi |0,\beta ,1)\\
& & \ \ \ \ \  T_{\epsilon}^{-1}(\frac{\xi}{2},\frac{\beta}
{i\pi },\frac{\beta}{i\pi},1,1|\frac{1+\xi}{2},\frac{1}{2},1,1|0,0,1)
T_{\epsilon}(\frac{1}{2},\frac{1+\xi}{2},\frac{\beta}
{i\pi },\frac{\beta}{i\pi},\xi |1,\frac{\xi}{2},1,1+\xi |2,0,1)\\
& &\stackrel{\sim}{F}_{\epsilon}(\beta)=\overline{\rho}_{\epsilon}^{'} 
g_{\epsilon}^{'\frac{1}{2}}(-2\beta)
T_{\epsilon}^{-1}(\frac{1}{2},2,\frac{1+\xi}{2},\frac{i\beta +\pi }{\pi },\xi |
1,\frac{\xi}{2},1,1+\xi |0,\beta ,1)\\
& & \ \ \ \ \  T_{\epsilon}^{-1}(\frac{\xi}{2},\frac{\beta}
{i\pi },\frac{\beta}{i\pi},1,1|\frac{1+\xi}{2},\frac{1}{2},1,1|0,0,1)
T_{\epsilon}(\frac{1}{2},\frac{1+\xi}{2},\frac{\beta}
{i\pi },\frac{\beta}{i\pi},\xi |1,\frac{\xi}{2},1,1+\xi |2,0,1)\\
& &\overline{G}_{\epsilon}(\beta)=\overline{g}_{\epsilon}(\beta)
T_{\epsilon}^{-2}(\frac{1+\xi}{2},\frac{\beta}
{2i\pi },\frac{\beta}{2i\pi},1,1|\frac{\xi}{2},\frac{1}{2},1,1|0,0,1)\\
& &G(\beta)=g_{\epsilon}(\beta)
T_{\epsilon}^{-2}(\frac{1}{2},\frac{1+\xi}{2},\frac{\beta}
{2i\pi },\frac{\beta}{2i\pi},1|\frac{\xi}{2},1,1,1|0,0,1)\\
& &\overline{G}_{\epsilon}^{'}(\beta)=\overline{g}_{\epsilon}^{'}(\beta)
T_{\epsilon}^{-2}(\frac{\xi}{2},\frac{\beta}
{2i\pi },\frac{\beta}{2i\pi},1,1|\frac{1+\xi}{2},\frac{1}{2},1,1|0,0,1)\\
& &W_{\epsilon}(\beta)=w_{\epsilon}(\beta)
T_{\epsilon}(\frac{1+\xi}{2},\frac{\beta}
{2i\pi },\frac{\beta}{2i\pi},1,1|\frac{\xi}{2},1,1,1|0,0,1)\\
& &\overline{H}_{\epsilon}(\beta)=\overline{h}_{\epsilon}(\beta)
T_{\epsilon}^{-2}(\frac{\beta}
{2i\pi },\frac{\beta}{2i\pi},1,1,1|\frac{1}{2},1,1,1|0,0,1)
\ \ ,\ \ U_{\epsilon}(\beta)=u_{\epsilon}(\beta)
T_{\epsilon}^{-2}(\frac{\beta}
{2i\pi },\frac{\beta}{2i\pi},1,1,1|1,1,1,1|0,0,1)\\
& &\overline{F}_{\epsilon}^{(0)}(\beta)=T_{\epsilon}^{2}
(\frac{2i\mu}{\pi}-1-\xi ,\frac{i\beta 
+\pi}{\pi},1,1,1|\xi ,1,1,1|0,\beta ,1)
T_{\epsilon}^{2}(\frac{3}{2},\frac{1-\xi }{2},\frac{i\beta 
+\pi}{\pi},1,1|\frac{\xi }{2},1,1,1|0,\beta ,1)\\
& &\ \ \ \ T_{\epsilon}^{2}(\frac{2i\mu}{\pi}-1-\xi ,\frac{\beta }{i\pi},
\frac{i\beta }{\pi},1,1|\xi ,2,1,1|1,0,1)
T_{\epsilon}^{2}(\frac{3}{2},\frac{1-\xi }{2},\frac{i\beta 
}{\pi},\frac{i\beta }{\pi},1|\frac{\xi }{2},2,1,1|1,0,1)\\
& &\ \ \ \ T_{\epsilon}^{-1}(\xi +\frac{2i\mu}{\pi},\frac{\beta}{i\pi},\frac{\beta}{i\pi},
1,1|\xi ,1,1,1|0,0,0)\\
& &F_{\epsilon}^{(0)}(\beta)=T_{\epsilon}^{-2}
(\frac{2i\mu}{\pi}-1-\xi ,\frac{i\beta 
+\pi}{\pi},1,1,1|\xi ,2,1,1|0,\beta ,1)
T_{\epsilon}^{-2}(\frac{3}{2},\frac{1-\xi }{2},\frac{i\beta 
+\pi}{\pi},1,1|\frac{\xi }{2},2,1,1|0,\beta ,1)\\
& &\ \ \ \ T_{\epsilon}^{-2}(\frac{2i\mu}{\pi}-1-\xi ,\frac{\beta }{i\pi},
\frac{i\beta }{\pi},1,1|\xi ,2,2,1|1,0,1)
T_{\epsilon}^{-2}(\frac{3}{2},\frac{1-\xi }{2},\frac{i\beta 
}{\pi},\frac{i\beta }{\pi},1|\frac{\xi }{2},2,2,1|1,0,1)\\
& &\ \ \ \ T_{\epsilon}(\xi +\frac{2i\mu}{\pi},\frac{\beta}{i\pi},\frac{\beta}{i\pi},
1,1|\xi ,2,1,1|0,0,0)\\
& &\overline{F}_{\epsilon}^{'(0)}(\beta)=T_{\epsilon}^{2}(\frac{2i\mu}{\pi}-1-\xi ,\frac{i\beta 
+\pi}{\pi},1,1,1|1+\xi ,1,1,1|0,\beta ,1)
T_{\epsilon}^{-2}(\frac{3}{2},\frac{1-\xi }{2},\xi,\frac{i\beta 
+\pi}{\pi},1|\frac{\xi }{2},1+\xi ,1,1|0,\beta ,1)\\
& &\ \ \ \ T_{\epsilon}^{-2}(\frac{2i\mu}{\pi}-1-\xi ,\frac{\beta }{i\pi},
\frac{i\beta }{\pi},1,1|1+\xi ,2,1,1|1,0,1)
T_{\epsilon}^{-2}(\frac{3}{2},\frac{1-\xi }{2},\xi ,\frac{i\beta 
}{\pi},\frac{i\beta }{\pi}|\frac{\xi }{2},2,1+\xi ,1|1,0,1)\\
& &\ \ \ \ T_{\epsilon}(\xi +\frac{2i\mu}{\pi},\frac{\beta}{i\pi},\frac{\beta}{i\pi},
1,1|1+\xi ,1,1,1|0,0,0)\\
& &\stackrel{\sim}{F}_{\epsilon}^{(0)}(\beta)=\{\overline{F}
_{\epsilon}^{'(0)}(\beta)\}^{-1} \ \ .
\end{eqnarray*}
Taking the limit of $\epsilon \longrightarrow 0$ ,one can find  
\begin{eqnarray*}
& &T_{\epsilon}(a_{1},a_{2},a_{3},a_{4},a_{5}|c_{1},c_{2},c_{3},c_{4}
|b_{1},b_{2},b_{3})\longrightarrow 
T(a_{1},a_{2},a_{3},a_{4},a_{5}|c_{1},c_{2},c_{3},c_{4}
|b_{1},b_{2},b_{3})\ \ ,\\
& &C_{\epsilon 1}\longrightarrow C_{1}\ \ ,\ \ 
C_{\epsilon 2}\longrightarrow C_{2}\ \ \ \ , 
\end{eqnarray*}
and some results which had been obtained by Lukyanov in Ref.[12]:
\begin{eqnarray*}
& &\rho_{\epsilon}\longrightarrow \rho \ \ ,\ \ 
\overline{\rho}_{\epsilon}\longrightarrow \overline{\rho} \ \ ,\ \ 
\overline{\rho}^{'}_{\epsilon}\longrightarrow \overline{\rho}^{'}\\
& &g_{\epsilon}(\beta) \longrightarrow g(\beta)\ \ ,\ \ 
\overline{g}_{\epsilon}(\beta) \longrightarrow \overline{g}(\beta)\ \ ,\ \ 
w_{\epsilon}(\beta) \longrightarrow w(\beta)\\ 
& & \overline{g}_{\epsilon}^{'}(\beta) \longrightarrow \overline{g}^{'}
(\beta)\ \ ,\ \ \overline{h}_{\epsilon}(\beta) \longrightarrow 
\overline{h}(\beta) \ \ ,\ \ \mu_{\epsilon}(\beta)\longrightarrow 
\mu (\beta)  \ \ .
\end{eqnarray*}
Therefore, we obtain function $
R_( \alpha_{1} .... \alpha_{k} | \delta_{1} ....\delta_{p} | \beta_{1}
 .... \beta_{n}|\gamma_{1}....\gamma_{r})$ as in (33).

\subsection*{A.2}
We here summarize formulas concerning coherent states of q-oscillator .

The coherent states $|z\rangle$ and $\langle \overline{z}|$ in the 
Fock spaces $H$  and its dual $\overline{H}$ of the q-oscillator 
$\{a_{n}\}$ in the section 4 ,are defined by 
\begin{eqnarray}
|z\rangle =exp\{\sum_{m=1}^{\infty} \frac{msh\frac{m\pi\epsilon\xi}{2}}
{sh\frac{m\pi\epsilon}{2}shm\pi\epsilon sh\frac{m\pi\epsilon(1+\xi)}{2}
}\ \ z_{n}a_{-n}\}|0\rangle \ \ \ \ ,\nonumber\\
\langle \overline{z}|=\langle \overline{z}|=\langle 0|
exp\{\sum_{m=1}^{\infty} \frac{msh\frac{m\pi\epsilon\xi}{2}}
{sh\frac{m\pi\epsilon}{2}shm\pi\epsilon sh\frac{m\pi\epsilon(1+\xi)}{2}
}\ \ \overline{z}_{n}a_{n}\}
\end{eqnarray}

The coherent states $\{|z\rangle\}$ (resp.$\{\langle \overline{z}|\}$ 
form a complete basis in the Fock space $H$ (resp. $\overline{H}$ ).
Namely it has the completeness relation 
\begin{eqnarray}
& &{\rm id}_{H}=\int \prod^{\infty}_{m=1}
\frac{msh\frac{m\pi\epsilon\xi}{2}dz_{m}d\overline{z}_{m}}
{sh\frac{m\pi\epsilon}{2}shm\pi\epsilon sh\frac{m\pi\epsilon(1+\xi)}{2}}
 e^{-\sum ^{\infty}_{m=1}
\frac{msh\frac{m\pi\epsilon\xi}{2}}
{sh\frac{m\pi\epsilon}{2}shm\pi\epsilon sh\frac{m\pi\epsilon(1+\xi)}{2}}
|z_{m}|^{2}}
|z_{m}\rangle \langle \overline{z}_{m}| \ \ \ .
\end{eqnarray}
Here the integration is taken over the entire complex plane .In the proof ,
the following integration formula is used:

\begin{eqnarray*}
& &\int \prod^{\infty}_{m=1}
\frac{msh\frac{m\pi\epsilon\xi}{2}dz_{m}d\overline{z}_{m}}
{sh\frac{m\pi\epsilon}{2}shm\pi\epsilon sh\frac{m\pi\epsilon(1+\xi)}{2}}
exp\{-\frac{1}{2}\sum ^{\infty}_{m=1}
\frac{msh\frac{m\pi\epsilon\xi}{2}}
{sh\frac{m\pi\epsilon}{2}shm\pi\epsilon sh\frac{m\pi\epsilon(1+\xi)}{2}}
(\overline{z}_{m},z_{m})A_{m}(\overline{z}_{m},z_{m})^{t}
+\sum^{\infty}_{m=1}(\overline{z}_{m},z_{m})B_{n}\}\\
& &\ \ \ =\prod^{\infty}_{m=1}(-detA_{n})^{\frac{1}{2}}
exp\{\frac{1}{2}\sum ^{\infty}_{m=1}\frac{sh\frac{m\pi\epsilon}{2}shm\pi\epsilon
sh\frac{m\pi\epsilon(1+\xi)}{2}}{msh\frac{m\pi\epsilon}{2}}B^{t}_{m}A^{-1}_{m}
B_{m}\} \ \ \ ,
\end{eqnarray*}
\begin{eqnarray}
\end{eqnarray}
where $A_{m}$ are invertible constant $2\times 2$ matrices and $B_{m}$ are 
constant 2 component vectors.

\end{document}